\PassOptionsToPackage{unicode}{hyperref}
\PassOptionsToPackage{hyphens}{url}
\documentclass[
]{article}
\usepackage[a4paper,margin=2.5cm]{geometry}
\usepackage{xcolor}
\usepackage{subcaption}
\usepackage{amsmath,amssymb}
\usepackage{iftex}
\ifPDFTeX
  \usepackage[T1]{fontenc}
  \usepackage[utf8]{inputenc}
  \usepackage{textcomp} 
\else 
  \usepackage{unicode-math} 
  \defaultfontfeatures{Scale=MatchLowercase}
  \defaultfontfeatures[\rmfamily]{Ligatures=TeX,Scale=1}
\fi
\usepackage{lmodern}
\ifPDFTeX\else
\fi
\IfFileExists{upquote.sty}{\usepackage{upquote}}{}
\IfFileExists{microtype.sty}{
  \usepackage[]{microtype}
  \UseMicrotypeSet[protrusion]{basicmath} 
}{}
\makeatletter
\@ifundefined{KOMAClassName}{
  \IfFileExists{parskip.sty}{%
    \usepackage{parskip}
  }{
    \setlength{\parindent}{0pt}
    \setlength{\parskip}{6pt plus 2pt minus 1pt}}
}{
  \KOMAoptions{parskip=half}}
\makeatother
\usepackage{longtable,booktabs,array}
\usepackage{calc} 
\usepackage{etoolbox}
\makeatletter
\patchcmd\longtable{\par}{\if@noskipsec\mbox{}\fi\par}{}{}
\makeatother
\IfFileExists{footnotehyper.sty}{\usepackage{footnotehyper}}{\usepackage{footnote}}
\makesavenoteenv{longtable}
\usepackage{graphicx}
\makeatletter
\newsavebox\pandoc@box
\newcommand*\pandocbounded[1]{
  \sbox\pandoc@box{#1}%
  \Gscale@div\@tempa{\textheight}{\dimexpr\ht\pandoc@box+\dp\pandoc@box\relax}%
  \Gscale@div\@tempb{\linewidth}{\wd\pandoc@box}%
  \ifdim\@tempb\p@<\@tempa\p@\let\@tempa\@tempb\fi
  \ifdim\@tempa\p@<\p@\scalebox{\@tempa}{\usebox\pandoc@box}%
  \else\usebox{\pandoc@box}%
  \fi%
}
\def\fps@figure{htbp}
\makeatother
\usepackage{svg}
\ifLuaTeX
  \usepackage{luacolor}
  \usepackage[soul]{lua-ul}
\else
  \usepackage{soul}
\fi
\usepackage{bookmark}
\IfFileExists{xurl.sty}{\usepackage{xurl}}{} 
\hypersetup{
  hidelinks,
  pdfcreator={LaTeX via pandoc}
}

\title{\textbf{Meta-PINNs: Meta-Learning Enhanced Physics-Informed Machine
Learning Framework for Turbomachinery Flow Predictions under Varying
Operation Conditions}}

\author{
Yuling Han, \quad Zhihui Li*, \quad Zhibin Yu \\[0.8em]
\textit{Department of Mechanical and Aerospace Engineering, School of Engineering,} \\
\textit{University of Liverpool, Liverpool, UK} \\[0.4em]
\texttt{\{yuling.han, zhihui.li, zhibin.yu\}@liverpool.ac.uk}
}

\date{}

\begin{document}

\maketitle

\textbf{ABSTRACT}

Coupling physics with machine learning models has shown great potential
for solving fluid dynamics problems governed by partial differential
equations. However, conventional methods, such as physics-informed
neural networks, often suffer from slow convergence, unstable training,
and limited generalization across different flow conditions. To overcome
these challenges, this study proposes a novel meta-learning enhanced
physics-informed neural networks (Meta-PINNs) framework, which
integrates a meta-optimization strategy into the training process. The
approach allows the model to automatically adapt its learning process to
varying physical regimes, thereby substantially improving both training
efficiency and predictive robustness. The proposed Meta-PINNs model is
evaluated on two representative flow problems: (1) unsteady flow around
a circular cylinder at multiple inlet Reynolds numbers, and (2) steady
turbulent flow within a compressor cascade passage at various angles of
attack. In both cases, the extrapolation performance of the developed
framework is comprehensively tested by predicting the flow fields at
Reynolds numbers and angles of attack that are not included in the
training set. The results demonstrate that Meta-PINNs achieve a 1-2
order-of-magnitude improvement in accuracy over vanilla physics-informed
neural networks and standard neural networks, while reducing
computational cost by up to 95.7 \% and 92.1 \%, respectively. It
successfully captures the sequential patterns of key flow features such
as pressure and velocity distributions under unseen conditions. Thus,
the findings confirm that the Meta-PINNs framework offers a notable
improvement in convergence and generalization over existing machine
learning approaches, providing a promising pathway toward smart
simulations of complex turbomachinery flows.


\vspace{0.5em}

\textbf{1 INTRODUCTION}

Cost-efficient prediction of turbomachinery flows in terms of
aerodynamic efficiency, pressure ratio and stability margin is central
to design optimization of modern gas turbines and aeroengines
\hyperref[_Ref214289851]{{[}1{]}}. Conventional computational fluid
dynamic (CFD) tools based on finite volume
\hyperref[_Ref214290103]{{[}2{]}} and finite elements
\hyperref[_Ref214290187]{{[}3{]}} approaches remain the primary way for
such analyses. However, the computational cost is prohibitive when
high-fidelity flow fields are required for refined turbomachinery
designs with geometric variations or when predictions must be generated
across a range of operating conditions, particularly during iterative
design stages.

Scientific machine learning (SciML) methods have emerged as powerful
surrogates for complex physical simulations by leveraging the rapid
development of advanced neural networks such as deep neural networks
\hyperref[_Ref214290356]{{[}4{]}}, convolutional neural networks
\hyperref[_Ref214290454]{{[}5{]}}, graph neural networks
\hyperref[_Ref214290483]{{[}6{]}}, Transformer
\hyperref[_Ref214290536]{{[}7{]}}, etc. Most of these surrogate models
are purely data-driven natures, where large, labeled datasets are
required to guide the model to learn the underlying patterns in hidden
spaces. However, the data-driven SciML models generally show poor
generalization performance outside the training distributions,
especially in off-design operating conditions. As a result, physically
inconsistent results are output from these models when extrapolating
\hyperref[_Ref214290986]{{[}8{]}}.

To mitigate the generalization issues from data-driven SciML models, a
straightforward way is to enforce the governing conservation physics
into the training process. Inspired by this, physics-informed neural
networks (PINNs) \hyperref[_Ref214291053]{{[}9{]}} have been proposed by
embedding the physical constraints in form of partial differential
equations (PDEs) in conjunction with boundary/initial conditions into
the loss terms. PINNs demonstrate unique advantages over conventional
CFD approaches for solving inverse problems, as they enable the
reconstruction of the entire flow field from sparse observational data,
which is particularly promising for digital twin applications. More
recently, PINNs also have been applied in turbomachinery flow
simulations {[}10-11{]}.

Although some advances achieved in PINNs-based frameworks, several
limitations are exposed in practical applications: 1) vanilla PINNs
always suffer from convergence issues especially in fluid simulation
scenarios due to the gradient pathology; 2) even converged, PINNs are
typically tailored to the specifically boundary condition setup, and
full retraining is necessary under varying boundary conditions. To
address these challenges, meta-learning or ``learning to learn''
strategy \hyperref[_Ref214291241]{{[}12{]}} is proposed as a
paradigm-shifting approach that enables rapid adaptation of PINNs to new
boundary conditions using only a small amount of additional data, i.e.,
few-shot learning. During a meta-learning phase, the meta-learning based
PINNs (Meta-PINNs) learn a set of shared parameters or latent
representations that serve as a good initialization for solving various
tasks. At meta-test phase, Meta-PINNs rapidly adapts to the previously
unseen operating conditions with sparse collation points within a few
training iterations. As a result, the physical flow fields can be
predicted at a significantly reduced computational cost.

The previous studies demonstrated that Meta-PINNs significantly reduced
training time and improved generalization performance when solving a
family of PDEs. Cheng and Alkhalifah \hyperref[_Ref214291251]{{[}13{]}}
demonstrated that the convergence speed and prediction accuracy were
significantly improved in estimating seismic wavefield when compared to
vanilla PINNs. Toloubidokhti et al. {[}14{]} further enhanced Meta-PINNs
by introducing adaptive loss weights and allocating optimal residual
points across different tasks. The superior performance of Meta-PINNs
compared to vanilla PINNs was validated through several benchmark tests.
Meta-PINNs have also been successfully applied to solve nonlinear
Schrodinger equation in previous study
\hyperref[_Ref214291261]{{[}15{]}}.

However, the majority of existing Meta-PINNs studies have been limited
to canonical PDE benchmarks, leaving a significant gap between
methodological research and real-world industrial applications,
especially for complex turbomachinery flow simulations. Motivated by
these challenges, we proposed a Meta-PINNs-based framework for complex
engineering flow simulations under varying conditions. As far as the
authors are aware, this is the first attempt to adapt Meta-PINNs for
turbomachinery flow simulations.

In this study, the fundamental methodologies of Meta-PINNs are shown in
Section 2. The Meta-PINNs-based framework is then developed and
benchmarked on two-dimensional (2D) cylinder flow cases with varying
inlet Reynolds numbers in Section 3. Subsequently, the benchmarked
framework is tailored for turbomachinery application where its
performance is evaluated on 2D compressor cascade flows under different
incoming attack angles in Section 4. Major conclusions are drawn at the
end.

\vspace{0.5em}

\textbf{2 NUMERICAL METHODS}

The methodologies employed in this work are detailed in this section,
together with the formulation of the proposed Meta-PINN framework and
its application to two representative flow configurations: the unsteady
laminar wake behind a circular cylinder under varying Reynolds numbers,
and the steady NACA-65 profile-based linear compressor cascade flow
under varying angles of attack. Benchmark assessments and the overall
dataset generation procedures are also provided to evaluate the
effectiveness of the adopted methods.

\textbf{2.1 Governing Equations}

The flow in both configurations is described by the incompressible
Navier--Stokes equations. For the compressor cascade case, the
incompressible assumption is justified as the flow corresponds to a
low-speed compressor cascade operating at a low Mach number
approximately 0.12. Under such conditions, compressibility effects are
negligible, and the pressure rise is primarily associated with flow
deceleration in accordance with Bernoulli's principle. In addition, the
numerical setup is designed to reproduce the experimental conditions
reported in the low-speed compressor cascade experiments conducted at
École Centrale de Lyon, further supporting the validity of the
incompressible flow assumption.

The continuity equation is:
\begin{equation}
\nabla \cdot \mathbf{u} = 0
\end{equation}

where \(\mathbf{u} = (u,v)\) is the velocity.

The general momentum equation is written as:
\begin{equation}
\frac{\partial \mathbf{u}}{\partial t}
+ \nabla \cdot (\mathbf{u} \otimes \mathbf{u})
= - \nabla p + \nabla \cdot \left( 2\nu_{\text{eff}} \mathbf{S} \right)
\end{equation}

where \(p\) is static pressure, and \(\nu_{eff}\) is the effective
viscosity. The strain--rate tensor \(\mathbf{S}\) is:

\begin{equation}
\mathbf{S} = \frac{1}{2}\left(\nabla \mathbf{u} + (\nabla \mathbf{u})^{T}\right)
\end{equation}

The two flow cases differ only in how \(\nu_{eff}\) is defined. For the
cylinder wake, the flow is laminar and the effective viscosity can be
written as:

\begin{equation}
\nu_{\text{eff}} = \nu
\end{equation}

where \(\nu\) represents the molecular viscosity.

For the compressor cascade flow, the flow is turbulent, and the
effective viscosity includes the eddy viscosity:

\begin{equation}
\nu_{\text{eff}} = \nu + \nu_t
\end{equation}

where \(\nu_{t}\) represents the turbulent viscosity from one-equation
Spalart-Allmaras (SA) \hyperref[_Ref215928477]{{[}16{]}} model.

\textbf{2.2 CFD Simulation and Dataset Preparation}

Before using the CFD solutions as training data for the Meta-PINNs, a
complete numerical simulation module is constructed using \emph{SU2}
\hyperref[_Ref215928589]{{[}17{]}} for each flow configuration.
\emph{SU2} is an open-source finite-volume-based CFD solver designed for
multiphysics analysis and aerodynamic shape optimization. In this study,
a velocity-pressure coupling scheme is used to update the flow velocity
and pressure simultaneously at each iteration. For unsteady simulations,
second-order dual-time stepping is employed for time marching. The
\emph{SU2} simulation stops when the maximum residual drops below
\(10^{- 6}\) or when the maximum number of iterations is reached. The
key design parameters and boundary conditions for each configuration are
listed in Tables 1 and 2.

\textbf{Table 1} \textbf{Reference values for cylinder flow}

{\def\LTcaptype{none} 
\begin{longtable}[]{@{}
  >{\centering\arraybackslash}p{(\linewidth - 2\tabcolsep) * \real{0.2627}}
  >{\centering\arraybackslash}p{(\linewidth - 2\tabcolsep) * \real{0.2118}}@{}}
\toprule\noalign{}
\begin{minipage}[b]{\linewidth}\centering
\textbf{Parameter}
\end{minipage} & \begin{minipage}[b]{\linewidth}\centering
\textbf{Value}
\end{minipage} \\
\midrule\noalign{}
\endhead
\bottomrule\noalign{}
\endlastfoot
Kinematic Viscosity & 1.00e-05 \(\frac{m^{2}}{s}\) \\
Density & 1 \(kg/m^{3}\) \\
Inlet Total Pressure & 0.02 \(Pa\) \\
Cylinder Diameter & 0.01 \(m\) \\
Domain Length & 0.3 \(m\) \\
Inlet Velocity Magnitude & {[}0.2 \(m/s\), 0.25 \(m/s\){]} \\
\end{longtable}
}

\textbf{Table 2 Reference values for compressor cascade flow}

{\def\LTcaptype{none} 
\begin{longtable}[]{@{}
  >{\centering\arraybackslash}p{(\linewidth - 2\tabcolsep) * \real{0.3153}}
  >{\centering\arraybackslash}p{(\linewidth - 2\tabcolsep) * \real{0.1592}}@{}}
\toprule\noalign{}
\begin{minipage}[b]{\linewidth}\centering
\textbf{Parameter}
\end{minipage} & \begin{minipage}[b]{\linewidth}\centering
\textbf{Value}
\end{minipage} \\
\midrule\noalign{}
\endhead
\bottomrule\noalign{}
\endlastfoot
Kinematic Viscosity & 1.51e-05 \(\frac{m^{2}}{s}\) \\
Density & 1.225 \(kg/m^{3}\) \\
Inlet Total Pressure & 980 \(Pa\) \\
Inlet Velocity Magnitude & 40 \(m/s\) \\
Chord length & 0.15 \(m\) \\
Design upstream flow angle & 54.31\({^\circ}\) \\
Design downstream flow angle & 31.09\({^\circ}\) \\
Aspect ratio & 2.47 \\
Solidity & 1.12 \\
\end{longtable}
}

For each flow case, the \emph{SU2} solutions and the Meta-PINNs
predictions are evaluated on the same computational mesh graph, allowing
for a direct pointwise comparison. Fig. 1(a) illustrates the
computational domain and mesh topology for the unsteady cylinder flow
with a velocity inlet and far-field atmospheric conditions. A no-slip
boundary condition is imposed on the cylinder surface. An H-type mesh is
employed in the wake region, while an O-type mesh is used upstream of
the cylinder. The total number of mesh cells is approximately 70,000.
For the compressor cascade case, an unstructured mesh is generated
within the passage, with local refinement near the blade surfaces. A
matched translational periodic boundary condition is applied on the
upper and lower boundaries to account for the influence of adjacent
blades. The blade is intentionally shifted toward the lower periodic
boundary to provide additional space for the flow separation developing
along the suction surface. The total mesh count for this case is
approximately 32,000.

\begin{figure}[htbp]
\centering
\includegraphics[width=0.8\textwidth]{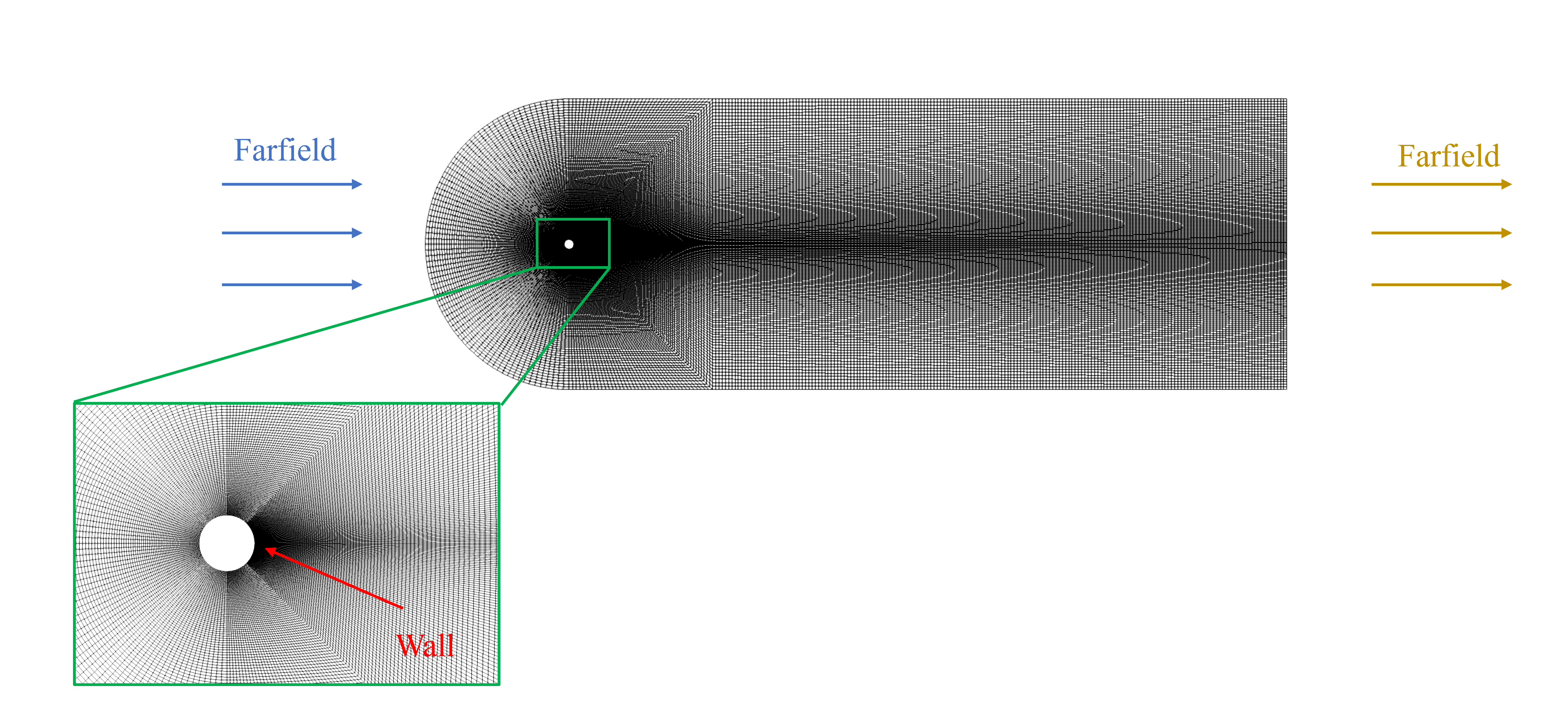}
\end{figure}

\begin{center}
(a)
\end{center}

\begin{figure}[htbp]
\centering
\includegraphics[width=0.8\textwidth]
{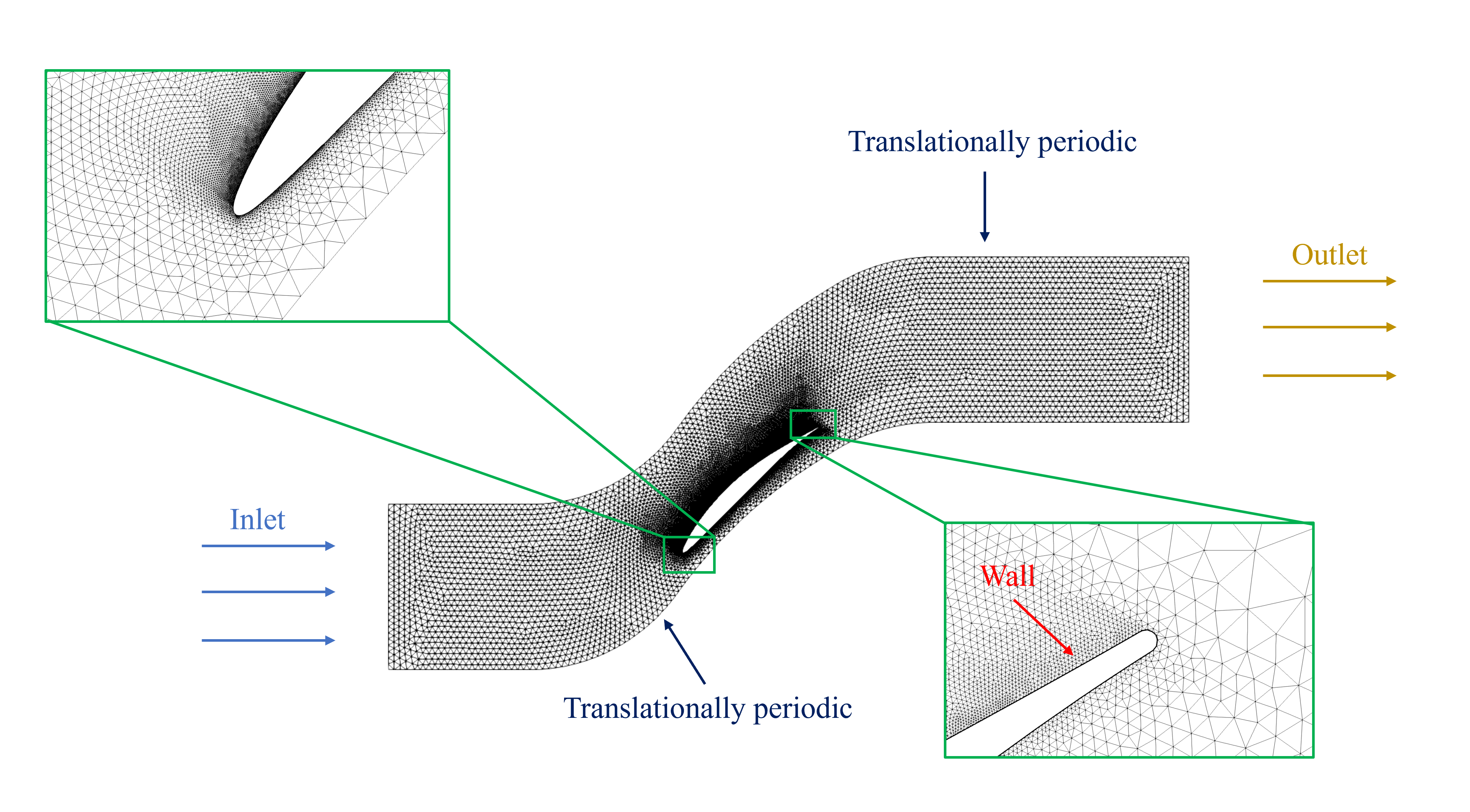}
\end{figure}

\begin{center}
(b)
\end{center}

\textbf{Fig. 1 (a) Computation domain and mesh topology for circular
cylinder flow case (b) Computation domain and mesh topology for
compressor cascade flow case}

For the cylinder flow case, the numerical simulation is first run for
several shedding periods, and snapshots of velocity and pressure are
extracted after the cylinder wake flow reaches a periodic pattern. For
the compressor cascade, the steady solutions for the velocity, pressure,
and effective viscosity fields are obtained. All fields are
postprocessed using \emph{Paraview} and serve as the datasets for
Meta-PINNs training, validation and testing. The accuracy of the
numerical setup is validated by comparing the simulation results with
available experimental measurements before using them as training
datasets.

\begin{figure}[htbp]
\centering

\begin{subfigure}{0.8\textwidth}
\centering
\includegraphics[width=\linewidth]{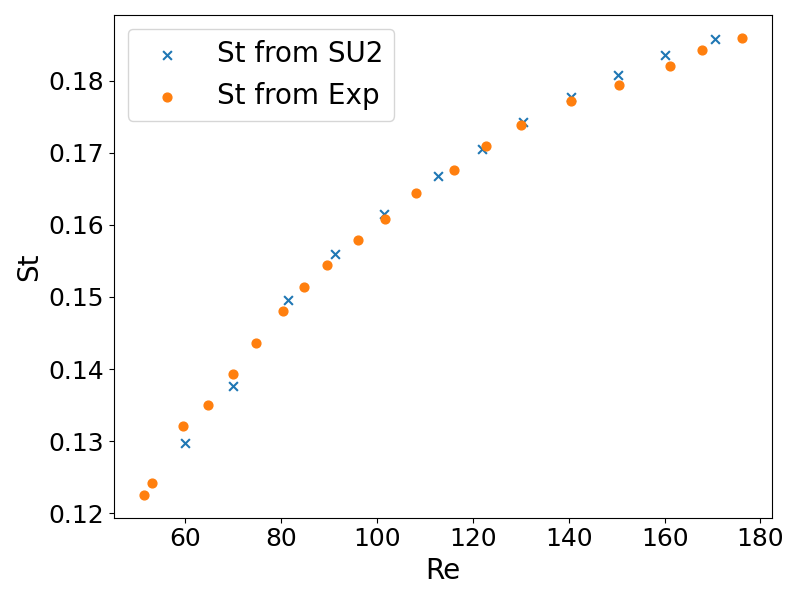}
\caption{}
\end{subfigure}

\vspace{0.5cm}

\begin{subfigure}{0.8\textwidth}
\centering
\includegraphics[width=\linewidth]{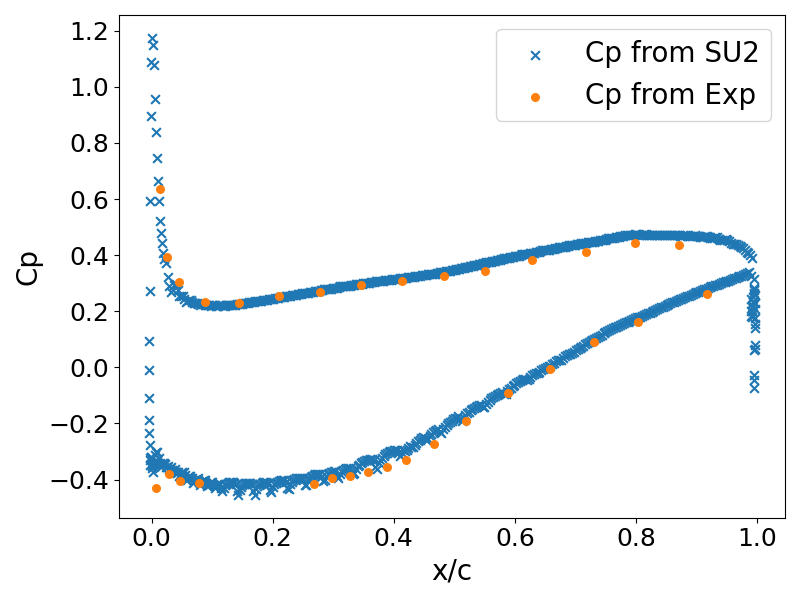}
\caption{}
\end{subfigure}

\begin{flushleft}
\textbf{Fig. 2} \textbf{(a) Strouhal number \emph{St}} \textbf{comparation between experiment data and CFD results (b) Static pressure coefficient \emph{Cp} comparation between experiment data and CFD results}
\end{flushleft}

\end{figure}

Figure 2(a) depicts the distribution of Strouhal number (\(St\)) as a
function of the inlet Reynolds number \(Re\). Strouhal number is a
dimensionless parameter characterizing the vortex-shedding frequency in
unsteady flows. It is defined as

\begin{equation}
S_t = \frac{fD}{U_{\infty}}
\end{equation}

where \(f\) is the vortex-shedding frequency, \(D\)is the characteristic
length, and \(U_{\infty}\)is the free stream velocity. The blue crosses
represent the simulation results from \emph{SU2} and the orange dots
mean the experimental data from Ref. \hyperref[_Ref215931448]{{[}18{]}}.
The Strouhal number generally increases with Reynolds number, indicating
that the vortex shedding frequency becomes higher at larger Reynolds
numbers. The simulation results well capture this \hl{}trend.

Figure 2(b) shows the comparison between the simulation results and
experimental data \hyperref[_Ref216015613]{{[}19{]}} for the cascade
pressure coefficient (\(C_{p}\)) along the chordwise direction at the
condition of the angle of attack \(\alpha = 0^{{^\circ}}\). The pressure
coefficient is a nondimensional measure of the pressure relative to the
free-stream static pressure, defined as

\begin{equation}
C_p = \frac{p - p_{\infty}}{\frac{1}{2}\rho U_{\infty}^{2}}
\end{equation}

where \(p\) is the local static pressure, \(p_{\infty}\) is the free
stream static pressure, \(\rho\) is the fluid density, and
\(U_{\infty}\) is the free stream velocity. The results demonstrate
excellent agreement between the predicted pressure coefficient and the
experimental measurements. These strong agreements confirm the
correctness of the numerical setup and indicate that the resulting
solutions serve as reliable reference datasets for training the machine
learning models.

Before being used for training and evaluation, all flow field physical
quantities are normalized. Spatial coordinates are normalized by the
characteristic length \(D\), velocity by the free-stream velocity
\(U_{\infty}\), pressure by \(\frac{1}{2}\rho U_{\infty}^{2}\), and
effective viscosity \hl{}\(\nu_{\text{eff}}\) by \(U_{\infty}D\). All
flow field results and error metrics reported in this study are based on
these normalized variables.

\textbf{2.3 Physics-Informed Neural Networks}

A fully connected neural network is generally employed to approximate
the mapping from spatial coordinates in conjunction with operation
condition parameters to target flow variables. For the cylinder flow
case, the network inputs are \((x,y,t,Re)\), which represent the
Cartesian coordinates, time, and Reynolds number, and the model outputs
are \((u,v,p)\), which represent the velocity components and static
pressure. For the compressor cascade case, the inputs are
\((x,y,\sin\alpha,\cos\alpha)\), which represent the Cartesian
coordinates and the trigonometric encoding of the angle of attack
\(\alpha\), and the outputs are \((u,v,p,\nu_{eff})\), which represent
the velocity components, static pressure, and effective turbulent
viscosity, respectively.

Automatic differentiation is used to evaluate spatial and temporal
derivatives in the PDE residuals. For a set of trainable parameters
\(\theta\), the physics-informed loss \(L_{phys}(\theta)\) is defined
as:

\begin{equation}
L_{\text{phys}}(\theta) =
\mathrm{MSE}\!\left(\mathcal{R}_{\text{mom}}\right)
+
\mathrm{MSE}\!\left(\mathcal{R}_{\text{cont}}\right)
\end{equation}

where \(MSE\) represents the mean squared error, and
\(\mathcal{R}_{\text{mom}}\) and \(\mathcal{R}_{\text{cont}}\) denote
the residuals of the momentum and continuity equations, respectively.
Supervised losses based on available data points \(L_{data(\theta)}\)
are incorporated as:

\begin{equation}
L_{\text{data}}(\theta)
=
\mathrm{MSE}(u_{\theta}, u)
+
\mathrm{MSE}(p_{\theta}, p)
+
\mathrm{MSE}(\nu_{\text{eff},\theta}, \nu_{\text{eff}})
\end{equation}

where \(u_{\theta}\), \(p_{\theta}\), \(\nu_{e_{ff},\theta}\) represents
the predicted velocity components, static pressure and effective eddy
viscosity, respectively. The overall PINN loss is then given by:

\begin{equation}
L_{\text{PINN}}(\theta)
=
\omega_d L_{\text{data}}(\theta)
+
\omega_p L_{\text{phys}}(\theta)
\end{equation}

where \(\omega_{\text{d}}\) and \(\omega_{\text{p}}\) denote the
adaptive weighting coefficients for each residual loss term, which are
typically annealed during training to stabilize convergence.

\textbf{2.4 Meta-Learning Framework}

To enable fast adaptation of the PINNs to previously unseen operating
conditions, a meta-learning strategy is adopted. Each case with a
varying Reynolds number or angles of attack defines a task
\(\mathcal{T}_{i}\), consisting of a support set
\(\mathcal{D}_{i}^{\text{sup}}\)for inner-loop adaptation, and a query
set \(\mathcal{D}_{i}^{\text{qry}}\)for meta-objective evaluation. Fig.
3 shows the architecture of the meta-learning PINN framework.

During meta-training, the model parameters \(\theta\) are first adapted
to each task after a small number of gradient steps:

\begin{equation}
\theta_{\dot{i}}' \leftarrow \theta - lr\nabla_{\theta}L_{PINN}\left( \theta;\mathcal{D}_{i}^{\sup} \right)
\end{equation}

where \(lr\) is the inner loop learning rate, \(\theta_{\dot{i}}'\)
represent the trained parameters for each case. The meta-objective
aggregates the performance of adapted parameters:

\begin{equation}
L_{meta(\theta)} = \sum_{i}^{}{L_{PINN}\left( \theta_{i}';\mathcal{D}_{i}^{qry} \right)}
\end{equation}

where \(L_{meta(\theta)}\) is the sum of PINNs residual loss for the
query set. The outer-loop update follows true meta-training:

\begin{equation}
\theta^{*} \leftarrow \theta - \beta\nabla_{\theta}L_{meta}(\theta)
\end{equation}

where \(\beta\) denotes the outer loop learning rate. This two-level
optimization yields a general purpose set of initial model parameters
\(\theta^{*}\) that can quickly adapt to new, unseen Reynolds numbers or
angles of attack with few additional data and physics constraints.

\begin{figure}[htbp]
\centering
\includegraphics[width=0.8\textwidth]{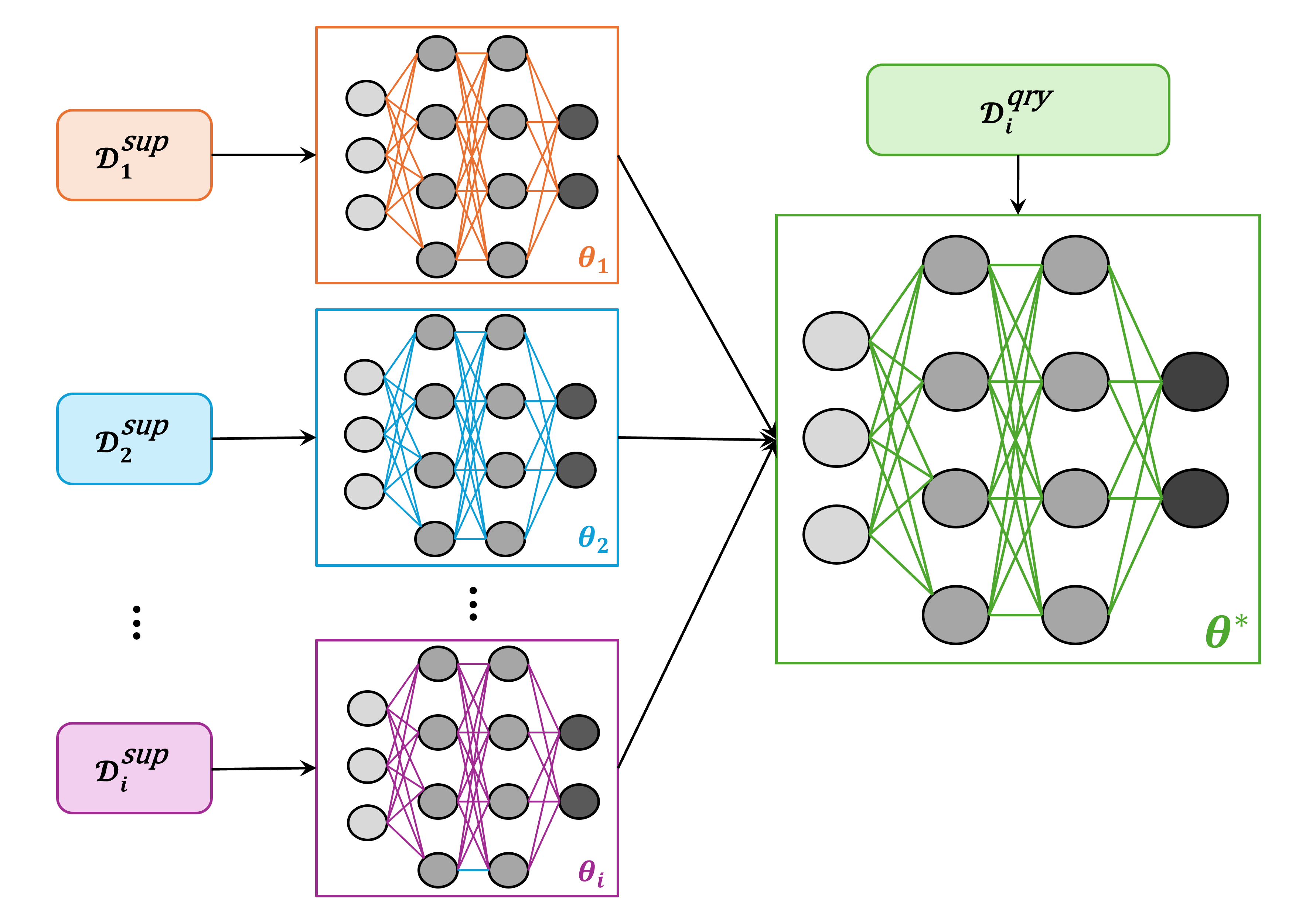}
\end{figure}
\begin{center}
\textbf{Fig. 3 Architecture of Meta-PINNs framework}
\end{center}

\textbf{2.5 Training Datasets and Task Construction}

For cylinder wake, flow-field snapshots at Reynolds numbers
\(Re = \{ 200,\ 210,\ 220,\ 230,\ 240,\ 250\}\) are generated using
\emph{SU2} solvers and serve as training dataset for Meta-PINNs. The
dataset contains velocity and pressure fields over 10 uniformly spaced
temporal snapshots per shedding cycle, covering 5 fully developed vortex
shedding periods, resulting in a total of 50 snapshots. The test tasks
correspond to the new conditions with\(Re = \{ 260,\ 300\}\),
representing extrapolative generalization. The training Reynolds numbers
are uniformly sampled within a moderate range to capture the main
vortex-shedding dynamics, while higher Reynolds numbers are reserved for
extrapolation to assess generalization beyond the training regime.

For the compressor cascade, two types of training tasks are constructed
to evaluate both interpolation and extrapolation performance. The
training angles of attack are
\(\alpha = \{ 0{^\circ},1{^\circ},2{^\circ},3{^\circ},4{^\circ},5{^\circ}\}\).
In the interpolation study, the model is tested on the angles of attack
of
\(\alpha = \{ 0.5{^\circ},1.5{^\circ},2.5{^\circ},3.5{^\circ},4.5{^\circ}\}\).
In the extrapolation study, the training tasks are chosen below the
target range, the model is tested on the angles of attack of
\(\alpha = \{ 6{^\circ},7{^\circ},8{^\circ},9{^\circ},10{^\circ}\}\).
This setup is designed to evaluate the ability of the model to
generalize to higher angles of attack, where flow separation and
nonlinear aerodynamic effects become more pronounced. The Meta-PINNs are
implemented in the \emph{PyTorch} environment, using 6 hidden layers
with 128 neurons in each layer. The trainable parameters in the model
are then fine-tuned using the \emph{Adam} optimizer
\hyperref[_Ref216030207]{{[}20{]}}.

\textbf{2.6 Computing Facility}

All training and testing procedures in this study were carried out on
the high-performance computing (HPC) platform \emph{Barkla} at the
University of Liverpool. The training processes of Meta-PINNs, PINNs and
NNs were conducted on one \emph{NVIDIA H100} GPU card. All tests were
performed using the same node configuration to ensure a fair comparison
of computational cost.

\vspace{0.5em}

\textbf{3 BENCHMARK TESTS ON CYLINDER FLOW}

To evaluate the generalization capability of the proposed Meta-PINNs
framework across different flow regimes, the unsteady laminar wake
behind a circular cylinder is first investigated. The Meta-PINNs are
trained using flow-field data corresponding to Reynolds numbers from 200
to 250. These cases represent moderate laminar vortex shedding
conditions with similar global structures but subtle variations in
shedding frequency and wake width. Figs. 4 and 5 present a comparison
between the Meta-PINNs predictions and the \emph{SU2} ground-truth
results for the velocity components and pressure contours at two new
Reynolds numbers, 260 and 300.

It can be observed that for Reynolds numbers of 260 and 300, the
Meta-PINNs accurately predict the unsteady velocity and pressure fields
at various sampling instants throughout a complete vortex shedding cycle
\(T\). The overall wake structure remains well preserved, and the
predicted velocity and pressure fields are highly consistent with the
CFD results across most regions. In the downstream area, the alternating
Kármán vortex street is clearly captured, and the relative locations and
strengths of the predicted vortex cores closely match those in the
ground-truth flow fields. These results demonstrate that Meta-PINNs
effectively learns the temporal evolution of the vortex street and
provides accurate flow field predictions throughout the entire vortex
shedding cycle.

The error distributions shown in Figs. 4 and 5 further highlight the
predictive characteristics of Meta-PINNs. In most regions of the flow
field, the pointwise relative errors remain small, consistent with the
overall agreement observed between the predicted and ground-truth
velocity and pressure fields. The largest errors are mainly concentrated
downstream the circular cylinder, where the velocity gradients are high
and the local flow structures vary rapidly. In these regions, slight
deviations in the predicted vortex position or strength can lead to
locally amplified errors. Despite these localized discrepancies, the
primary wake features are accurately captured. This further confirms
that Meta-PINNs are capable of learning the dominant wake dynamics, with
only minor differences appearing in the most unstable regions.

The relative error fields also clearly reveal the distinct error
patterns at different Reynolds numbers. In the \(Re = 260\) test case,
the prediction errors of both the velocity and pressure are noticeably
smaller than those for case with \(Re = 300\). This is because the case
at \(Re = 260\) lies closer to the Reynolds number range used for
training, and its wake dynamics more closely resemble the flow
characteristics present in the training samples. Since the Reynolds
number of 300 lies well outside the training range, the Meta-PINNs
performs a larger step extrapolation, resulting in increased errors in
certain regions, particularly within the highly unstable vortex cores.
Nevertheless, the primary wake structures are still successfully
captured.

To further quantify the prediction accuracy beyond the visualized
results, the root-mean-square error(RMSE) of the velocity and pressure
fields for the different test Reynolds numbers are compared. The RMSE
analysis enables a more detailed assessment of the model performance and
reveals how the prediction error varies as the flow complexity
increases. By comparing the results at \(Re\) of 260 and 300, the
influence of the test \(Re\) on the generalization capability of
Meta-PINNs becomes more clearly evident.

\begin{figure}[htbp]
\centering

\begin{subfigure}{0.8\textwidth}
\centering
\includegraphics[width=\linewidth]{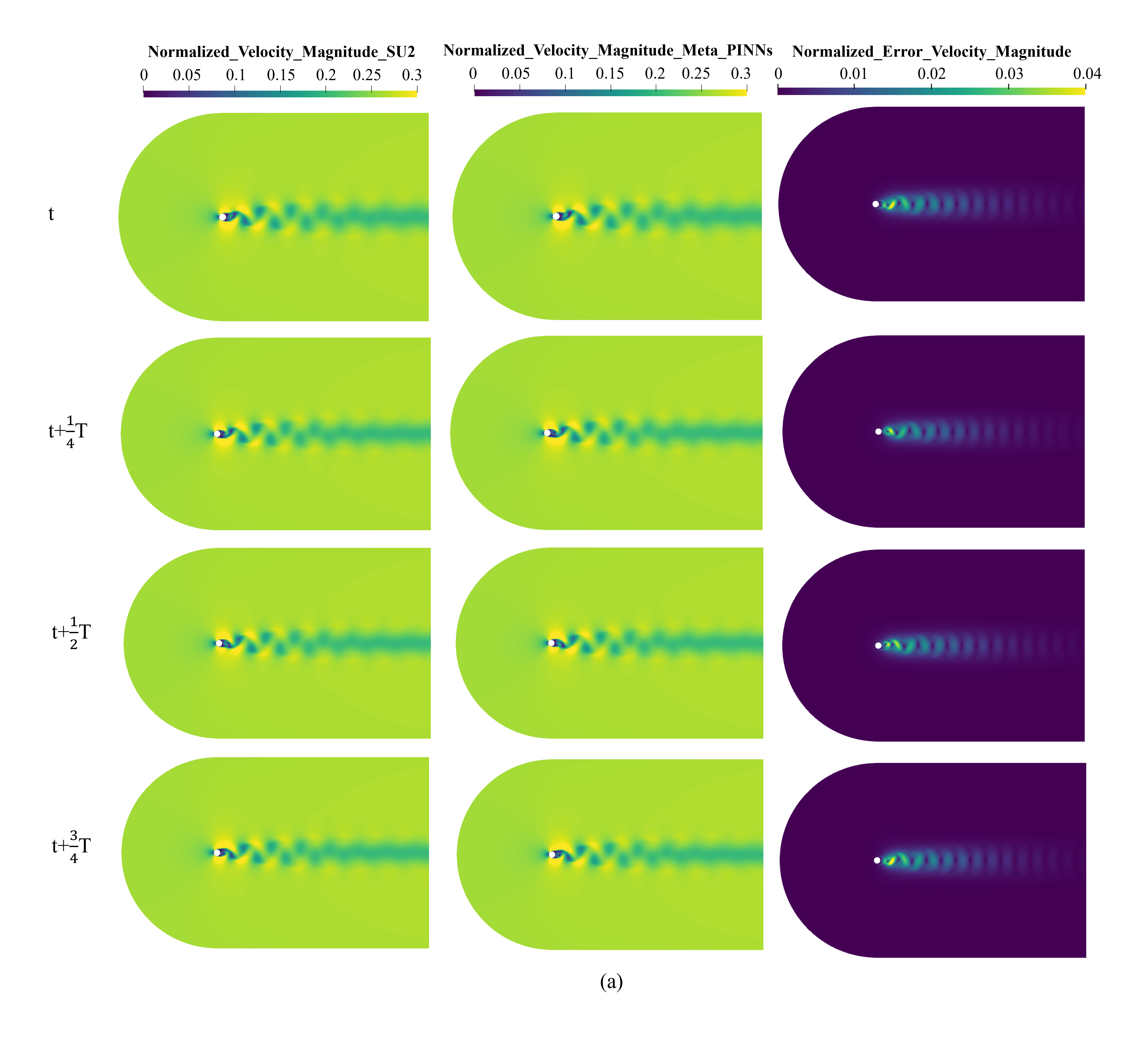}
\end{subfigure}

\vspace{0.5cm}

\begin{subfigure}{0.8\textwidth}
\centering
\includegraphics[width=\linewidth]{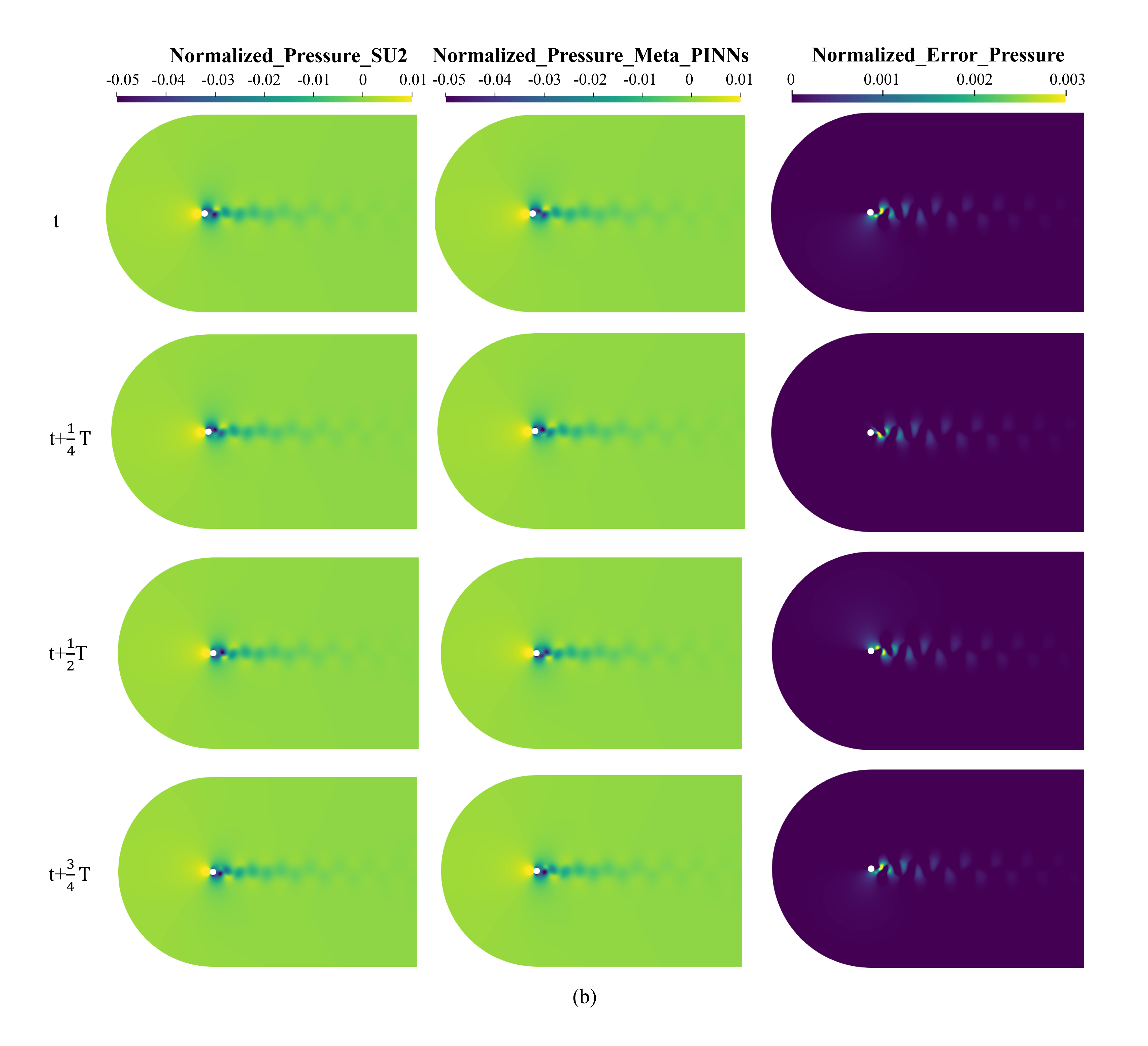}
\end{subfigure}

\begin{flushleft}
\textbf{Fig. 4 Comparison of the normalized velocity magnitude and
pressure fields between CFD and Meta-PINNs at different instants within
one shedding cycle for Reynolds number 260}
\end{flushleft}

\end{figure}

\begin{figure}[htbp]
\centering

\begin{subfigure}{0.8\textwidth}
\centering
\includegraphics[width=\linewidth]{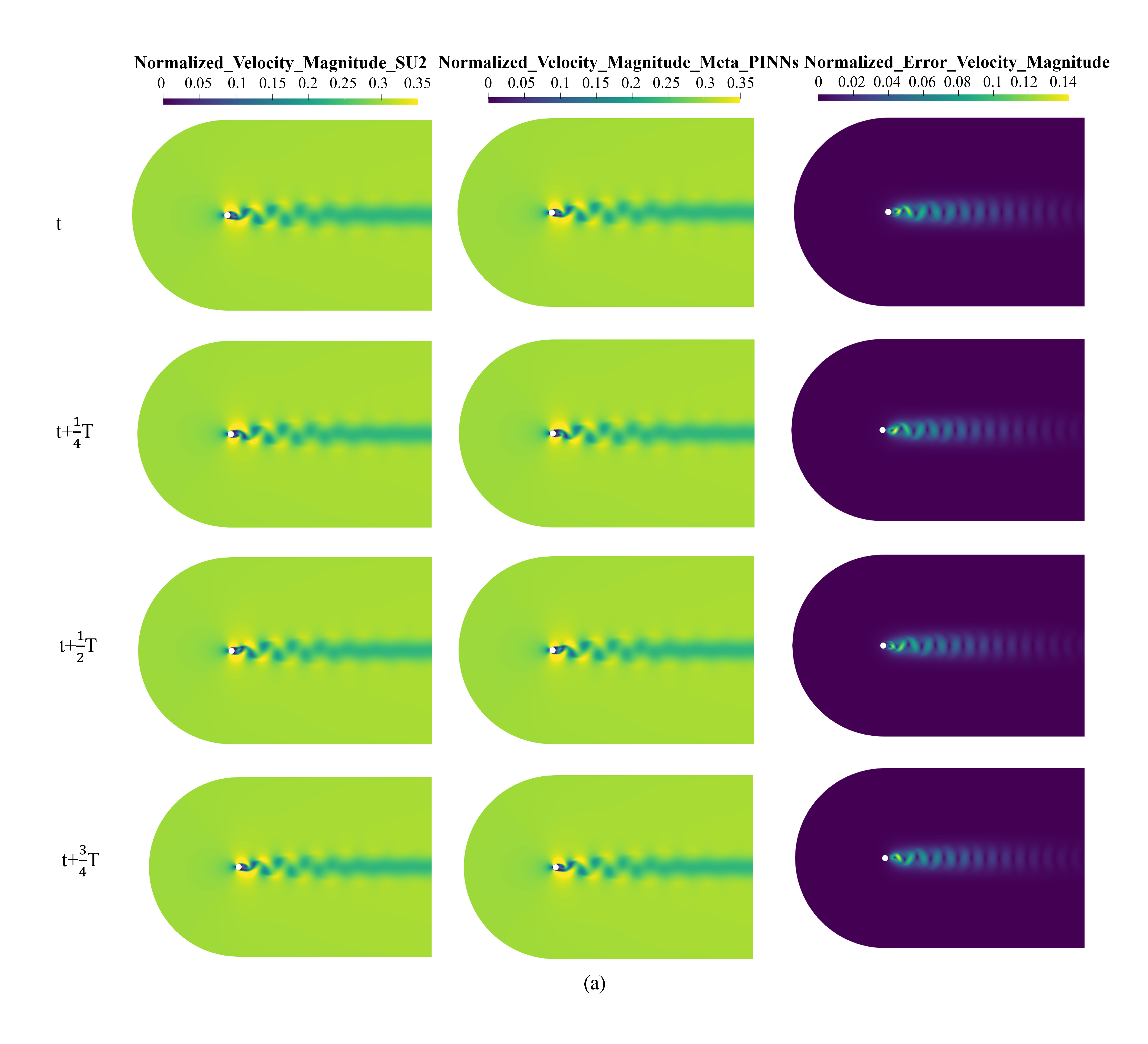}

\end{subfigure}

\vspace{0.5cm}

\begin{subfigure}{0.8\textwidth}
\centering
\includegraphics[width=\linewidth]{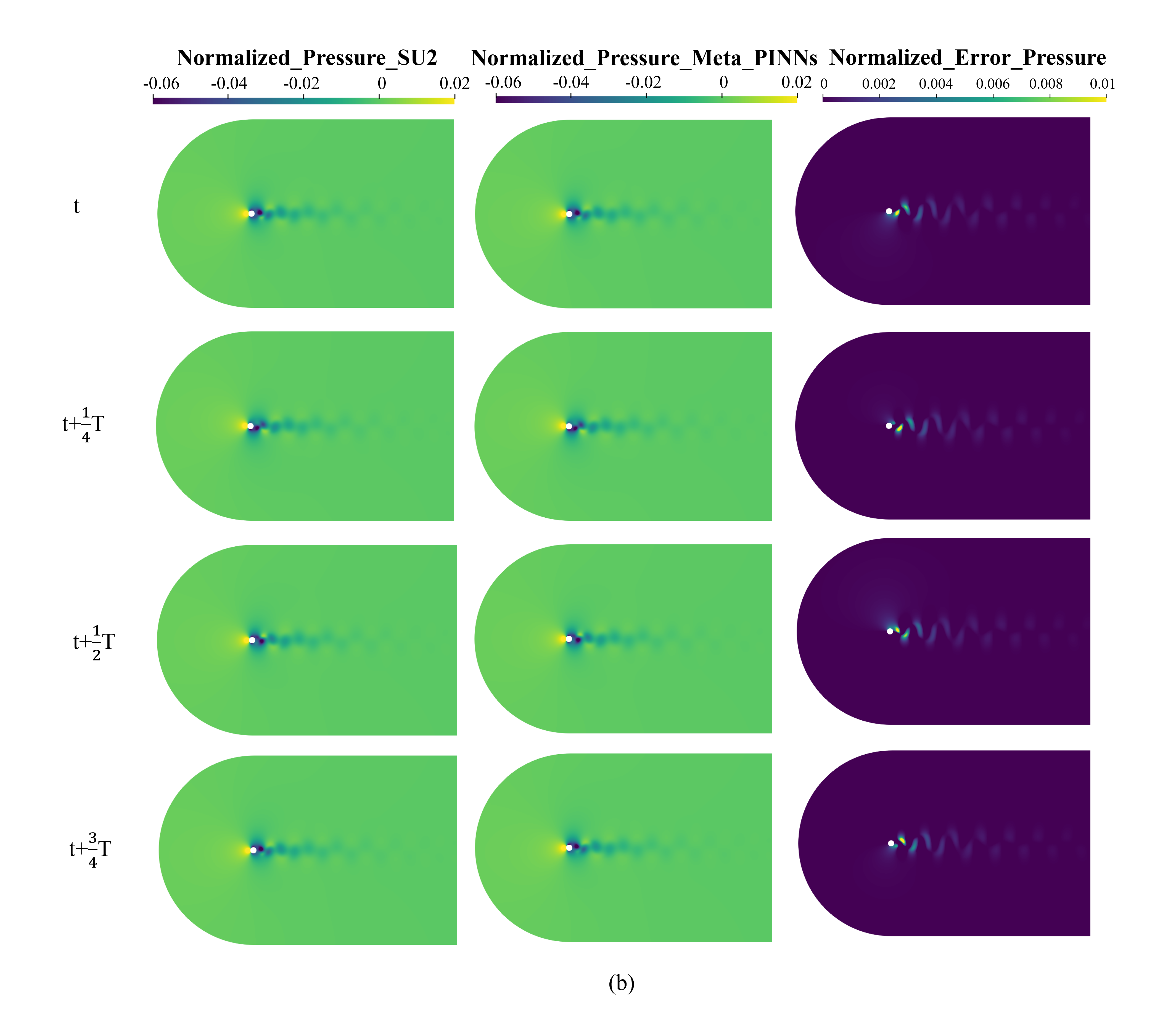}

\end{subfigure}

\begin{flushleft}
\textbf{Fig. 5 Comparison of the normalized velocity magnitude and
pressure fields between CFD and Meta-PINNs at different instants within
one shedding cycle for Reynolds number 300}
\end{flushleft}

\end{figure}

As shown in Tab. 3, both the velocity and pressure fields exhibit
relatively small RMSE values at \(Re\  = \ 260\), indicating a low level
of overall prediction error. In contrast, all predicted physical
quantities show a clear increase in RMSE at \(Re\  = \ 300\). This
increase in error at \(Re = 300\) is consistent with the qualitative
comparison shown in Figs. 4 and 5. Since extrapolation is inherently
more prone to error, the RMSE levels are significantly higher than those
at \(Re\  = \ 260\). To validate the effectiveness of the proposed
method, Meta-PINNs performance is compared to the conventional neural
networks (NNs) and the standard PINNs in terms of computational cost and
prediction accuracy. To ensure a fair comparison, the Meta-PINNs, PINNs,
and NNs all employ the same neural network architecture, consisting of
six hidden layers with 128 neurons per layer and identical activation
functions, such that performance differences are solely attributed to
the learning strategy.

\textbf{Table 3 RMSE comparison of Meta-PINNs predictions at Reynolds
numbers 260 and 300 based on normalized flow-field variables}

{\def\LTcaptype{none} 
\begin{longtable}[]{@{}
  >{\centering\arraybackslash}p{(\linewidth - 4\tabcolsep) * \real{0.1841}}
  >{\centering\arraybackslash}p{(\linewidth - 4\tabcolsep) * \real{0.1446}}
  >{\centering\arraybackslash}p{(\linewidth - 4\tabcolsep) * \real{0.1471}}@{}}
\toprule\noalign{}
\begin{minipage}[b]{\linewidth}\centering
\textbf{Re}
\end{minipage} & \begin{minipage}[b]{\linewidth}\centering
\textbf{260}
\end{minipage} & \begin{minipage}[b]{\linewidth}\centering
\textbf{300}
\end{minipage} \\
\midrule\noalign{}
\endhead
\bottomrule\noalign{}
\endlastfoot
RMSE\_\(u\) & 0.004 & 0.014 \\
RMSE\_\emph{v} & 0.006 & 0.022 \\
RMSE\_\(V_{m}\) & 0.007 & 0.027 \\
RMSE\_\(p\) & 0.001 & 0.002 \\
\end{longtable}
}

\begin{figure}[htbp]
\centering
\includegraphics[width=0.8\textwidth]{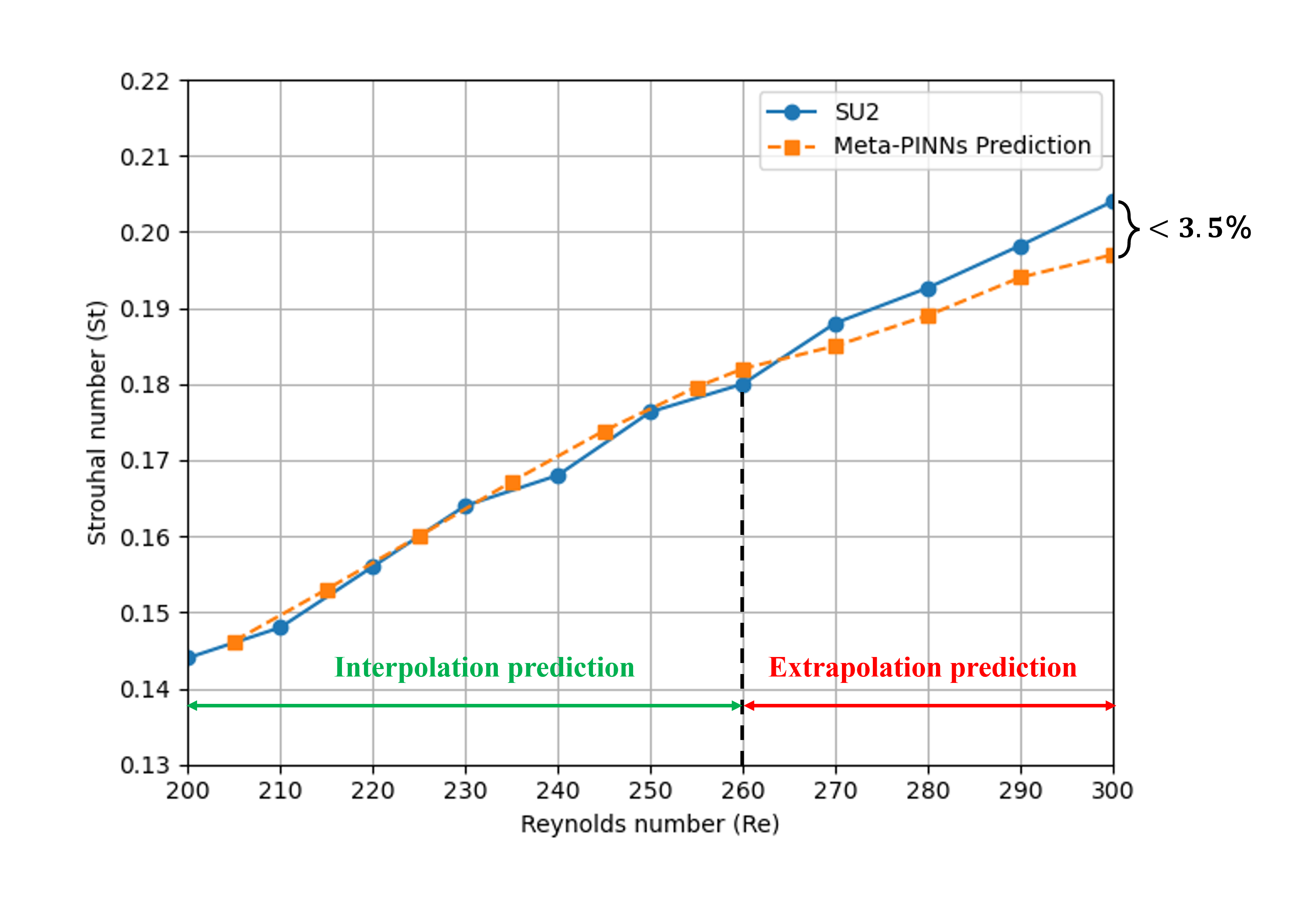}

\textbf{Fig. 6 Strouhal number comparison between CFD results and predictions from Meta-PINNs}

\label{fig:strouhal}
\end{figure}

The Strouhal numbers predicted in the Reynolds number range of 200 to
300 are compared with those obtained from CFD simulations, as shown in
Fig. 6. Within this range, the Strouhal number increases gradually with
Reynolds number, and the Meta-PINNs successfully capture this trend.
When the Reynolds number is close to the training data regime, the
predicted values align closely with the \emph{SU2} reference results,
exhibiting only small errors. As the Reynolds number increases further,
Meta-PINNs begin to underestimate the shedding frequency, and the
deviation grows, reaching a maximum error smaller than 3.5\% at
\(Re\  = \ 300\). The increasing error with Reynolds number can be
attributed to the stronger Kármán vortex shedding and the heightened
sensitivity of wake dynamics at higher \(Re\). Even so, the predicted
Strouhal numbers still correctly characterize the overall trend across
the entire test interval.

\textbf{Table 4 Computational cost and prediction accuracy comparison of
Meta-PINNs, PINNs and NNs at Re = 260 based on normalized flow-field
variables}

{\def\LTcaptype{none} 
\begin{longtable}[]{@{}
  >{\centering\arraybackslash}p{(\linewidth - 6\tabcolsep) * \real{0.1053}}
  >{\centering\arraybackslash}p{(\linewidth - 6\tabcolsep) * \real{0.1317}}
  >{\centering\arraybackslash}p{(\linewidth - 6\tabcolsep) * \real{0.1185}}
  >{\centering\arraybackslash}p{(\linewidth - 6\tabcolsep) * \real{0.1209}}@{}}
\toprule\noalign{}
\begin{minipage}[b]{\linewidth}\centering
\textbf{Re}
\end{minipage} & \begin{minipage}[b]{\linewidth}\centering
\textbf{Meta-PINNs}
\end{minipage} & \begin{minipage}[b]{\linewidth}\centering
\textbf{PINNs}
\end{minipage} & \begin{minipage}[b]{\linewidth}\centering
\textbf{NNs}
\end{minipage} \\
\midrule\noalign{}
\endhead
\bottomrule\noalign{}
\endlastfoot
RMSE\_\(u\) & \textbf{0.004} & 0.244 & 0.321 \\
RMSE\_\emph{v} & \textbf{0.006} & 0.373 & 0.533 \\
RMSE\_\(V_{m}\) & \textbf{0.007} & 0.255 & 0.327 \\
RMSE\_\(p\) & \textbf{0.001} & 0.034 & 0.438 \\
Time cost & \textbf{7:59:33} & 7d-18:43:20 & 4d-5:24:09 \\
\end{longtable}
}

In Table 4, the PINNs are implemented as standard physics-informed
neural networks by enforcing the governing equations and boundary
conditions through physics-based loss terms, while the NNs serve as
purely data-driven baselines without physical constraints. The
prediction accuracy is evaluated using RMSE computed on normalized
velocity and pressure fields. The computational cost reported
corresponds to the total wall-clock time required for model training and
prediction.

To validate the effectiveness of the proposed method, Meta-PINNs
performance is compared to the conventional neural networks (NNs) and
the standard PINNs in terms of computational cost and prediction
accuracy. To ensure a fair comparison, the neural architectures in the
Meta-PINNs, PINNs, and NNs are maintained consistently throughout this
study. As shown in Table 4, Meta-PINNs outperform both PINNs and NNs in
terms of prediction accuracy and training efficiency. In detail, based
on the RMSE results, the prediction accuracy of the Meta-PINNs improves
by at least 1 to 2 orders of magnitude compared with the PINNs and NNs.
Meanwhile, the computational cost is reduced by approximately 95.7 \%
and 92.1 \% relative to the PINNs and NNs, respectively. These results
demonstrate that the Meta-PINNs are both more accurate and more
efficient for multi-condition flow prediction.

The improved performance of Meta-PINNs over standard PINNs can be
attributed to the meta-learning strategy. By learning shared
representations across multiple flow conditions during the meta-training
phase, Meta-PINNs provide a favorable initialization for new tasks.
Consequently, the subsequent training becomes more task-adaptive and
converges more efficiently than PINNs trained from scratch, leading to
both higher accuracy and lower computational cost.

\vspace{0.5em}

\textbf{4 APPLICATION ON COMPRESSOR FLOW}

The compressor cascade case uses training data from angles of attack
between \(0^{\circ}\)and \(5^{\circ}\). The Meta-PINNs first predict the
flow fields at the intermediate angles of attack of \({0.5}^{\circ}\),
\({1.5}^{\circ}\), \({2.5}^{\circ}\), \({3.5}^{\circ}\)and
\({4.5}^{\circ}\). Fig. 7(a) shows the velocity vector comparison from
\(\alpha = {0.5}^{\circ}\) to \(\alpha = {4.5}^{\circ}\), including the
\emph{SU2} reference fields, the Meta-PINNs predictions and the
corresponding errors. The predicted flow field matches the reference
solution well across the entire computational domain. The local
acceleration in the front region of the suction side is accurately
reproduced. The flow direction and streamline curvature within the
passage remain consistent with the ground-truth field, and the wake
downstream of the trailing edge exhibits the correct strength and
thickness. The prediction error remains small across most of the domain.
Relatively large local errors are mainly observed in the regions near
the inlet and outlet, where the flow is strongly influenced by boundary
conditions and undergoes rapid global acceleration or deceleration.
These errors primarily originate from boundary-condition-induced global
velocity gradients. Although the errors in these regions are larger than
those in other parts of the domain, they remain confined to limited
areas and do not affect the overall flow structure. This indicates that
the Meta-PINNs are able to preserve the dominant topological features of
the flow field within the interpolation range and achieve accurate
full-field velocity predictions.

\begin{figure}[tbp]
\centering

\includegraphics[width=0.8\textwidth]{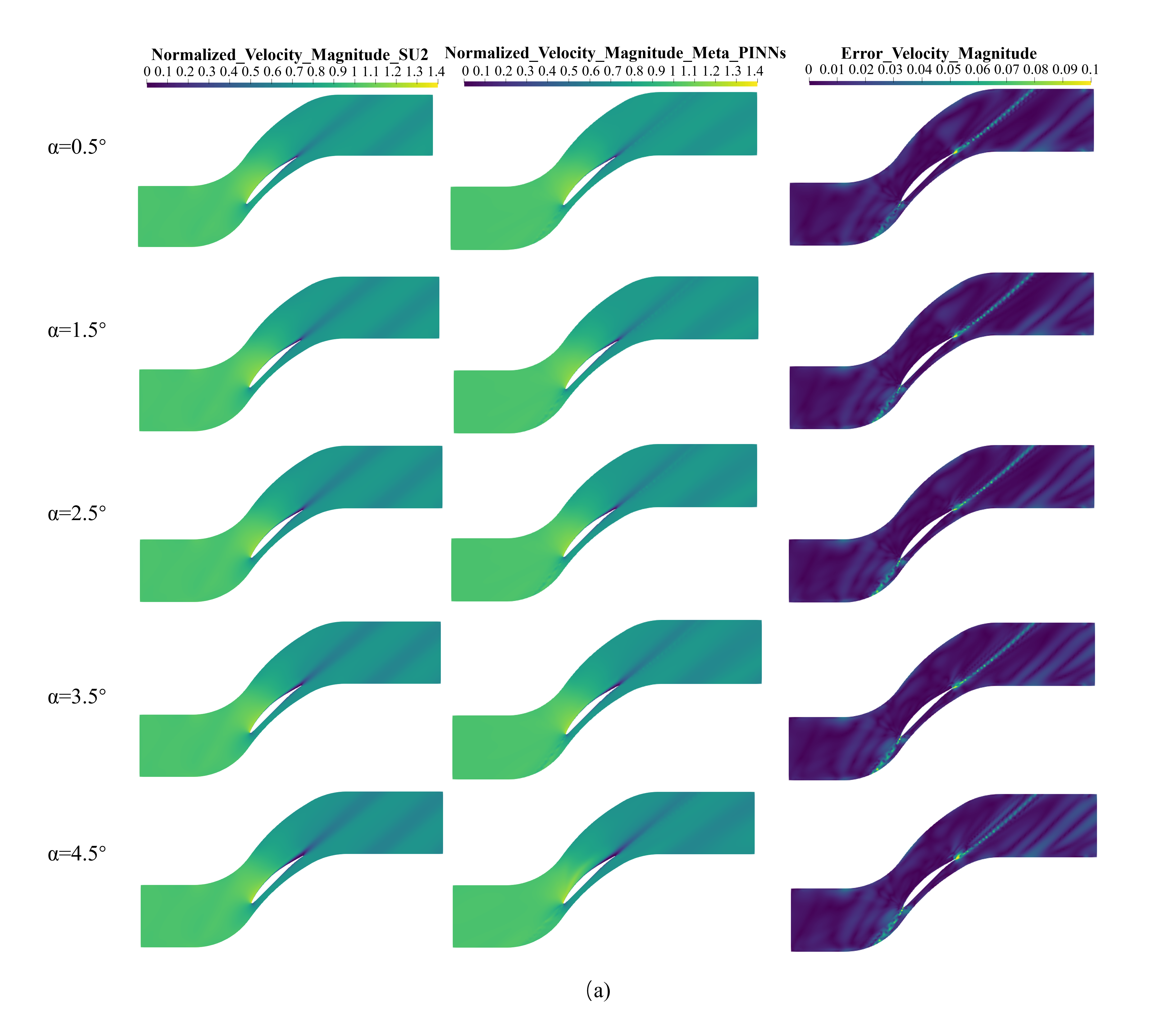}

\vspace{0.4cm}

\includegraphics[width=0.8\textwidth]{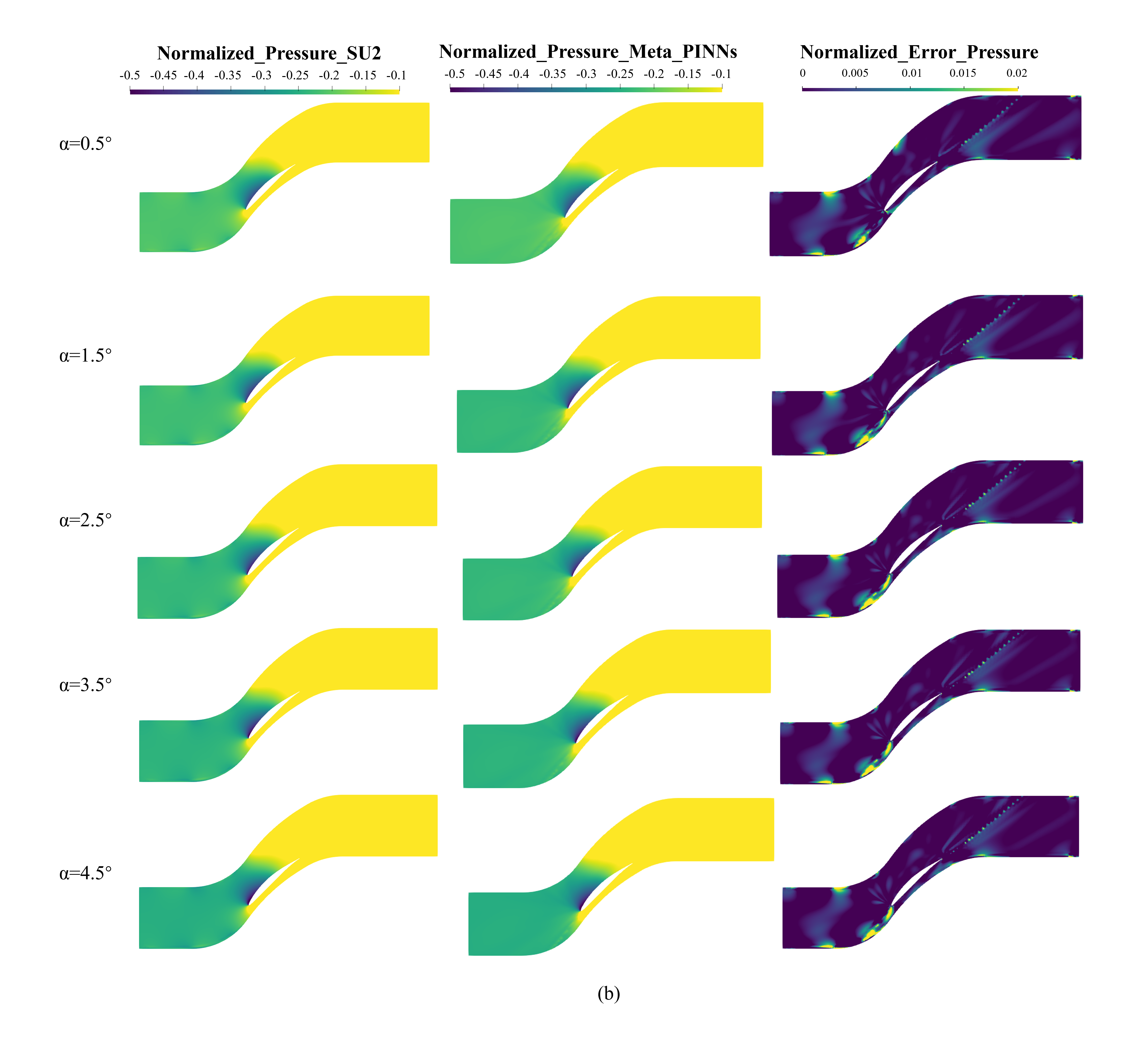}

\end{figure}

\begin{figure}[!t]
\centering
\includegraphics[width=0.8\textwidth]{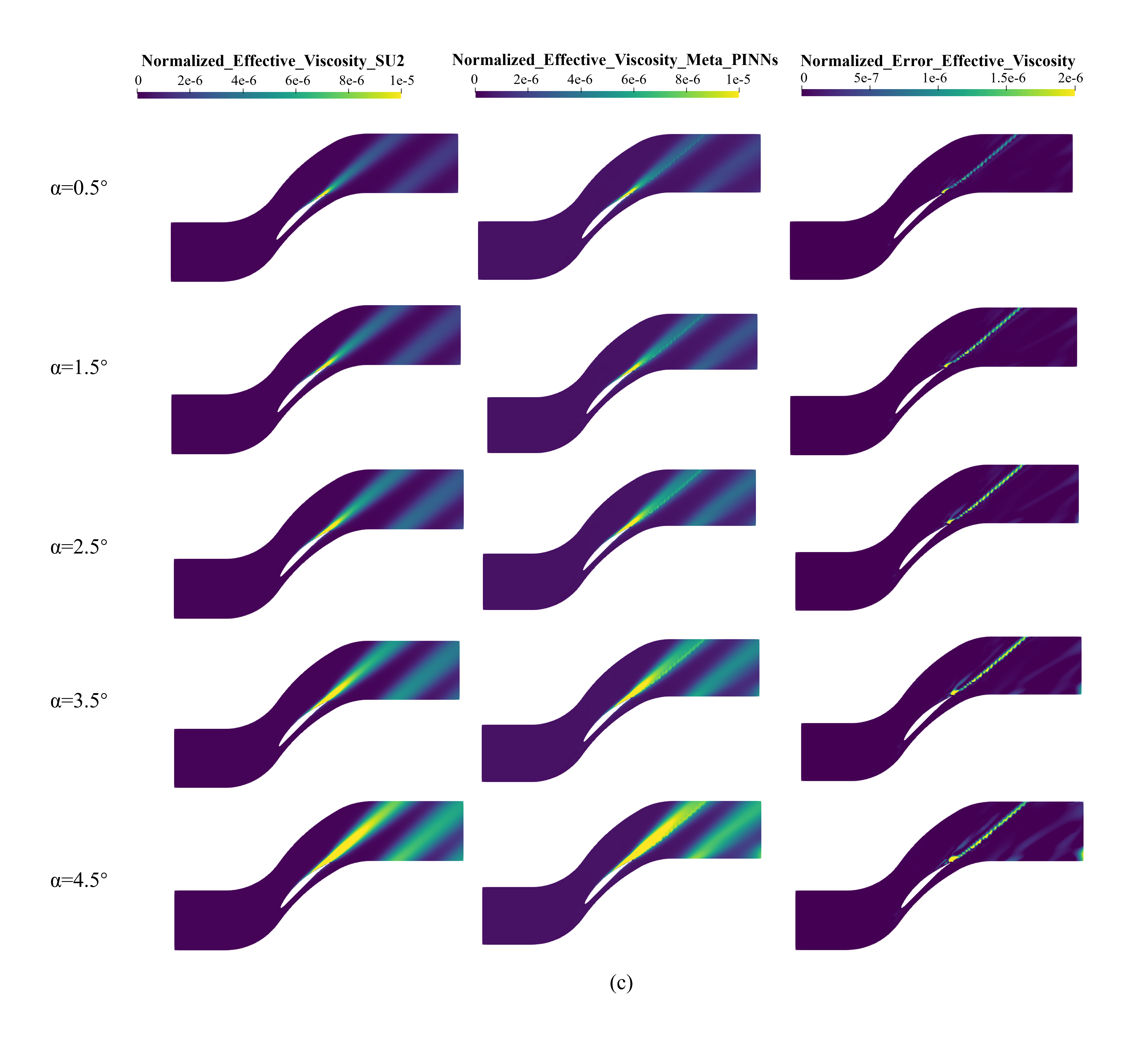}

\begin{flushleft}
\textbf{Fig. 7 Comparison of normalized velocity magnitude, pressure,
and effective viscosity fields predicted by CFD and Meta-PINNs for
angles of attack
from }\(\mathbf{0.5}^{\mathbf{{^\circ}}}\mathbf{\ }\)\textbf{to }
\(\mathbf{4.5}^{\mathbf{{^\circ}}}\)
\end{flushleft}

\end{figure}

From Fig. 7(b), it can be seen that Meta-PINNs also provides accurate
predictions of the pressure field, with the overall pressure
distribution closely matching the reference results. The low pressure
region on the suction side is correctly captured, and the magnitude of
the pressure rise on the pressure side is also accurately predicted. The
pattern of the pressure distribution within the channel, as well as the
direction of the pressure gradients, is consistent with the ground-truth
solution. The pressure prediction error remains small over most of the
computational domain, with relatively larger errors primarily
concentrated near the inlet and outlet regions, as well as in localized
zones where the flow undergoes rapid deformation. These regions are
characterized by strong pressure gradients induced by boundary
conditions and flow acceleration or deceleration, which makes the
pressure field more sensitive to minor prediction deviations. Although
the local errors in these areas are comparatively higher, they are
confined to limited regions and do not distort the overall pressure
distribution. This demonstrates that, under interpolation conditions,
Meta-PINNs are capable of accurately capturing the global pressure field
while preserving its dominant structural characteristics.

Fig. 7(c) presents a comparison of the effective viscosity fields in the
blade cascade case, including the SU2 reference solution, the Meta-PINNs
prediction, and the corresponding error distribution. The effective
viscosity accounts for both molecular viscosity and turbulent eddy
viscosity effects, and is highly sensitive to the generation and
transport of turbulence. Overall, the effective viscosity predicted by
the Meta-PINNs shows good agreement with the SU2 reference solution
within the blade passage. The streamwise distributions and magnitude
levels of effective viscosity along both the suction and pressure
surfaces are accurately reconstructed, indicating that the model is
capable of effectively learning the turbulence intensity characteristics
inside the cascade channel. The prediction error of effective viscosity
remains small over most of the computational domain and is mainly
concentrated in the downstream wake region behind the trailing edge,
where a distinct high error band can be observed. In this region, the
effective viscosity varies rapidly in space, making the prediction more
sensitive to local deviations. Despite the localized increase in error,
the discrepancies remain confined to the wake region and do not disrupt
the overall structure of the effective viscosity field. This
demonstrates that, under interpolation conditions, the Meta-PINNs can
reliably reconstruct the effective viscosity distribution.

\begin{figure}[tbp]
\centering

\includegraphics[width=0.8\textwidth]{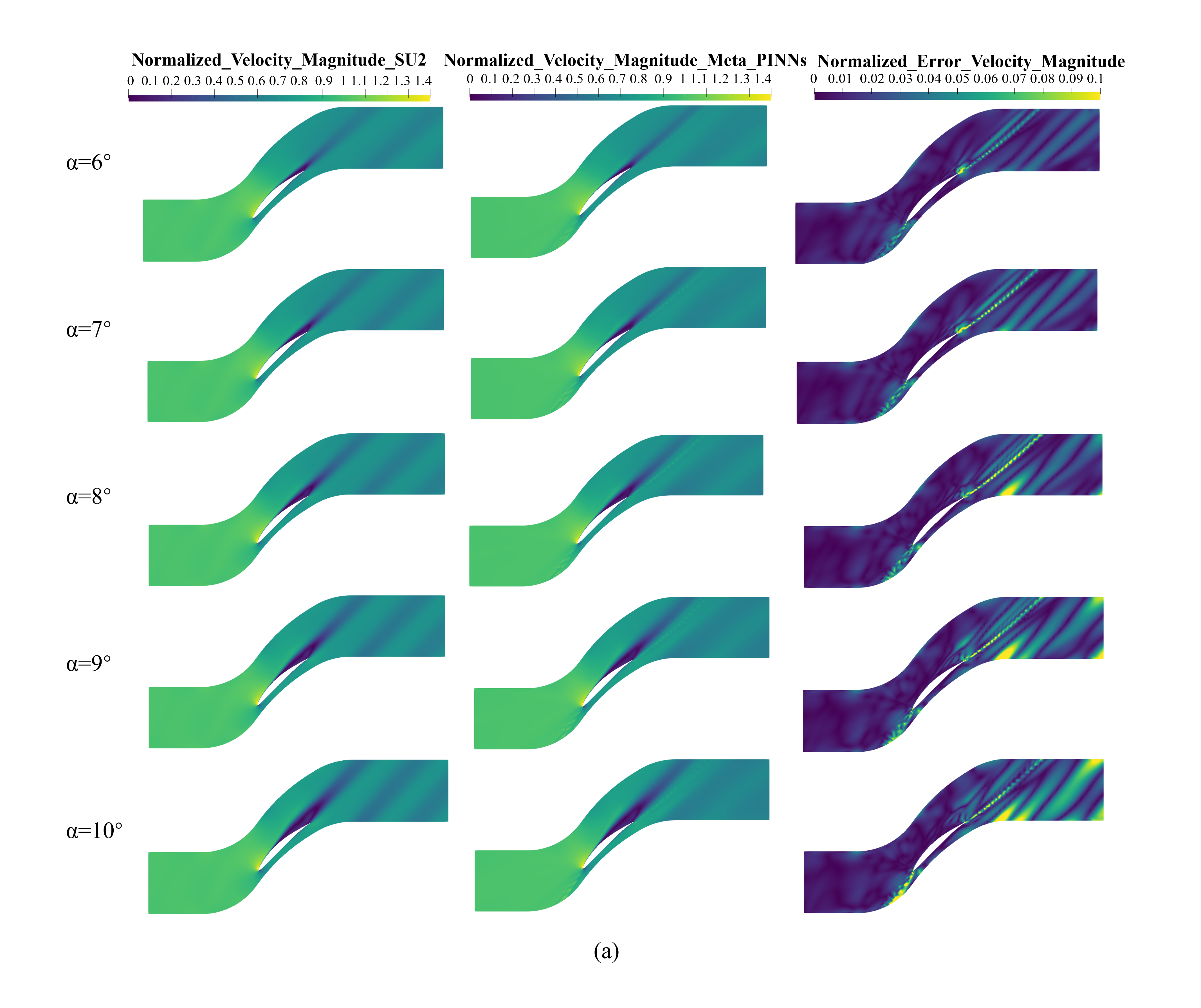}

\vspace{0.4cm}

\includegraphics[width=0.8\textwidth]{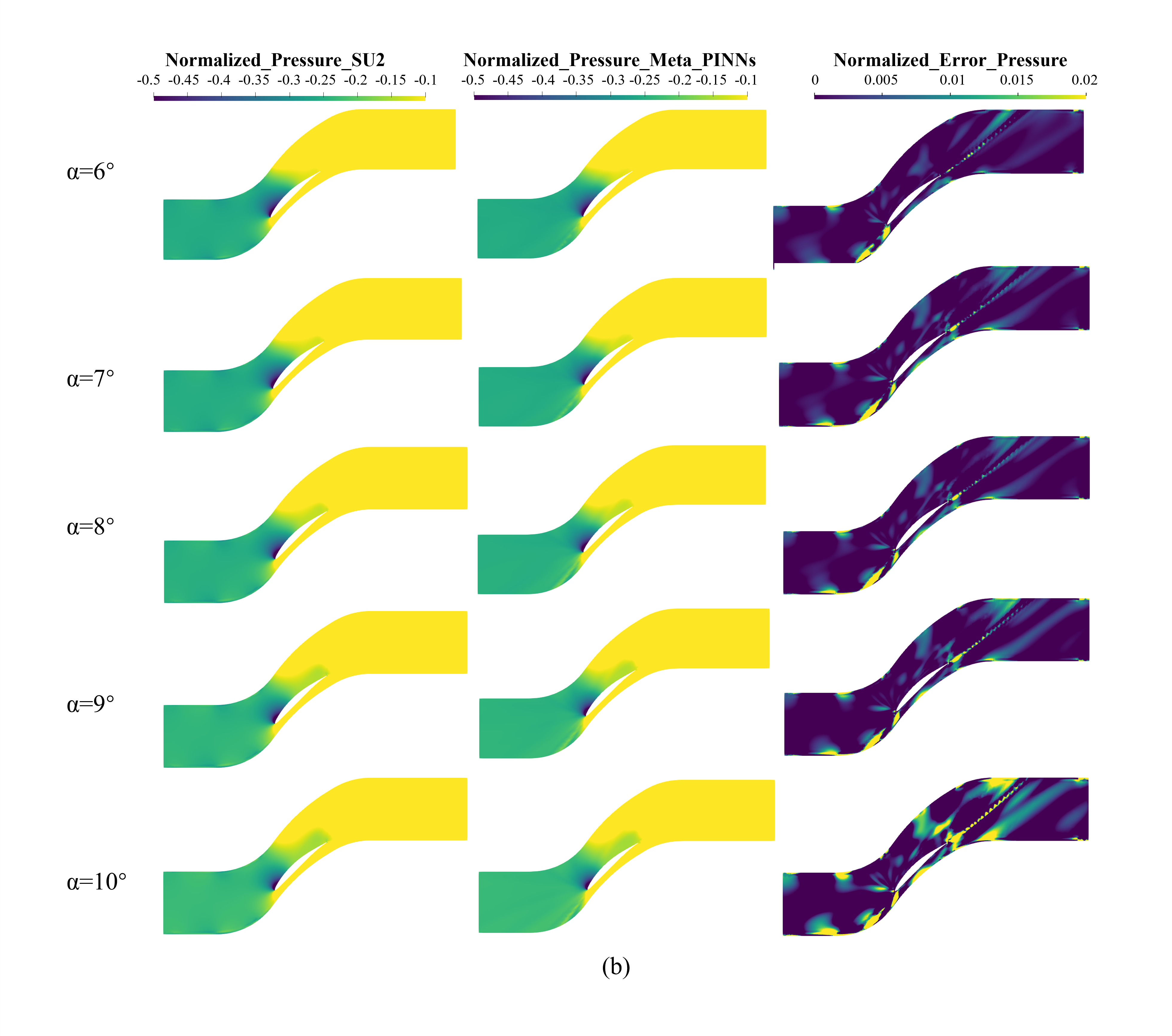}

\end{figure}

\begin{figure}[tbp]
\centering
\includegraphics[width=0.8\textwidth]{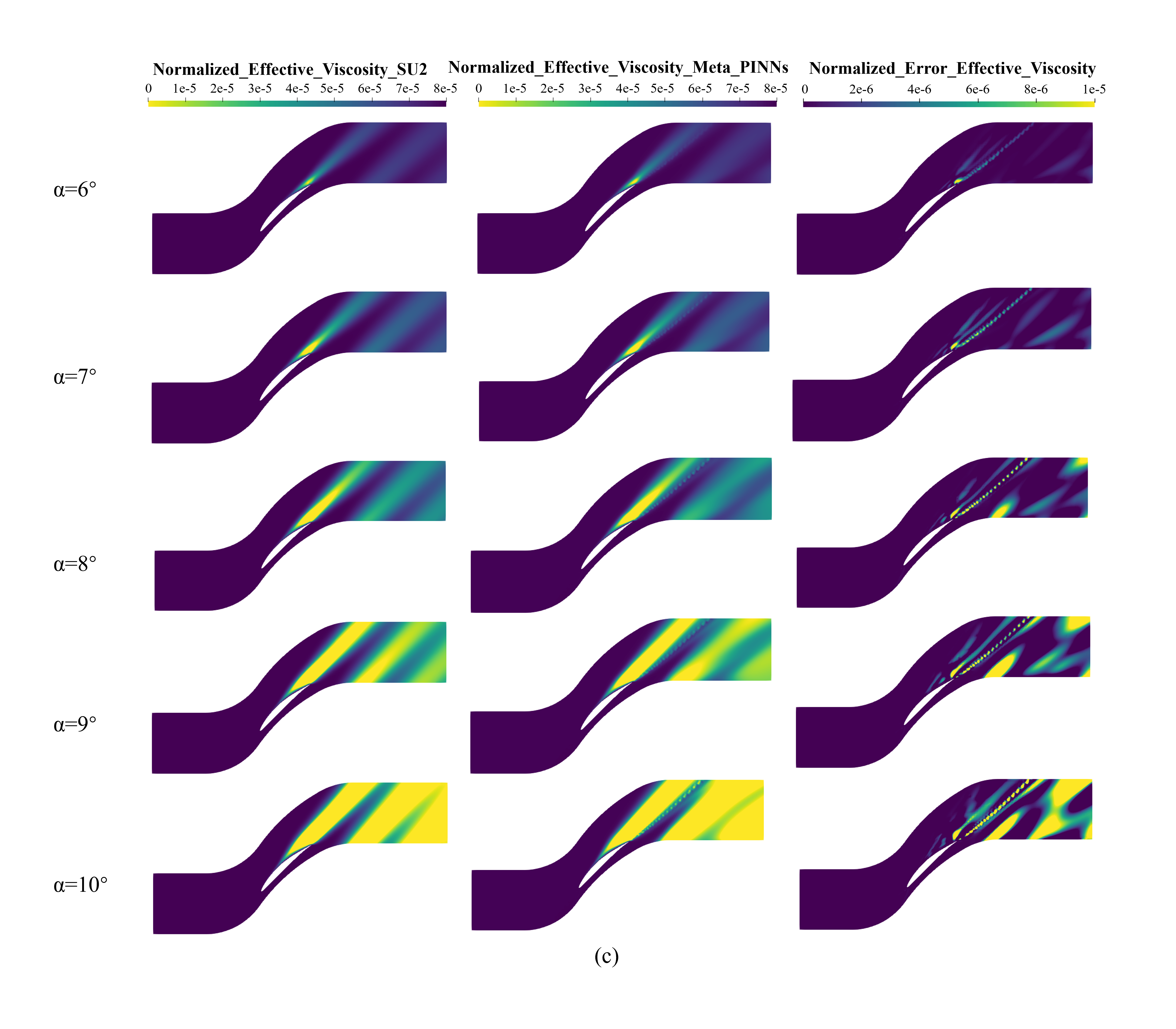}

\begin{flushleft}
\textbf{Fig. 8 Comparison of normalized velocity magnitude, pressure,
and effective viscosity fields predicted by CFD and Meta-PINNs for
angles of attack from }\(\mathbf{6}^{\mathbf{{^\circ}}}\mathbf{\ }\)\textbf{to }
\(\mathbf{10}^{\mathbf{{^\circ}}}\)

\end{flushleft}
\end{figure}

In the extrapolation tests, the prediction range of Meta-PINNs extends
to the angles of attack between 6° and 10°, as shown in Fig. 8. The
prediction results show that Meta-PINNs are still able to reconstruct
the overall flow structure with reasonable accuracy. The acceleration
region on the suction side and the flow-direction variations inside the
channel follow the correct trends, and the wake pattern downstream of
the trailing edge remains largely consistent with the reference
solution.

Figure 8(a) presents the predicted velocity fields and the corresponding
error distributions under extrapolation conditions with angles of attack
ranging from 6° to 10°. Overall, the velocity fields predicted by the
Meta-PINNs show good agreement with the SU2 reference solutions, and the
dominant flow structures are consistently preserved across all operating
conditions. Within the blade passage, including both the suction and
pressure side regions, the prediction errors remain relatively small,
indicating that the primary flow features learned from the training data
can be effectively generalized beyond the interpolation range. More
noticeable discrepancies are mainly observed near the inlet and outlet
regions, where the flow is strongly influenced by boundary conditions
and global flow adjustment processes. As the angle of attack further
increases, the error distribution downstream of the trailing edge
gradually exhibits more distinct structural patterns, with errors
predominantly concentrated in the wake region. These discrepancies
remain spatially localized and do not significantly affect the overall
velocity distribution or the global flow topology. This suggests that
the Meta-PINNs retain a robust capability for full-field velocity
prediction even under extrapolation conditions.

Figure 8(b) illustrates the pressure error distributions under
extrapolation conditions with angles of attack ranging from 6° to 10°.
Overall, the pressure fields predicted by the Meta-PINNs show good
agreement with the SU2 reference solutions, and the global pressure
distribution within the blade passage is consistently preserved across
all operating conditions. For all extrapolation cases, relatively larger
pressure errors are observed near the inlet and outlet regions. This
behavior is expected, as the pressure field is a globally coupled
quantity that is strongly constrained by boundary conditions and
influenced by the overall flow adjustment process. Consequently, small
prediction deviations tend to be amplified near the domain boundaries.
In addition to the inlet and outlet regions, another error feature
becomes increasingly pronounced with increasing angle of attack:
band-shaped error patterns emerge and intensify near the lateral
boundaries of the passage. These regions are strongly affected by the
periodic boundary matching of the blade cascade, and their sensitivity
to changes in operating conditions increases under high-angle-of-attack
scenarios. Despite the localized growth of errors with increasing angle
of attack, the dominant pressure topology and the overall loading trends
within the passage remain well preserved. This indicates that the
Meta-PINNs maintain robust pressure field prediction capability even
under extrapolation conditions.

Fig. 8(c) presents the error distributions of the effective viscosity
under extrapolation conditions with angles of attack ranging from 6° to
10°. Overall, the effective viscosity predicted by the Meta-PINNs
remains in good agreement with the SU2 reference solutions, and the main
distribution characteristics inside the blade passage are consistently
preserved. Across all extrapolation cases, the prediction error remains
small over most of the computational domain. Relatively larger errors
are primarily observed near the outlet and in the wake region downstream
of the trailing edge. As the angle of attack increases, the error
magnitude in these regions gradually increases, while the error
distribution remains spatially localized. The increased errors near the
outlet and in the wake are associated with strong turbulence production
and rapid spatial variations of effective viscosity in these regions,
which makes the prediction more sensitive under extrapolation
conditions. Despite the localized error amplification at higher angles
of attack, the overall structure of the effective viscosity field
remains intact, indicating that the Meta-PINNs can reliably reconstruct
effective viscosity distributions even beyond the training range.

\textbf{Table 5 RMSE comparison of Meta-PINNs predictions at different
angles of attack, RMSE values are computed based on normalized
flow-field variables}

{\def\LTcaptype{none} 
\begin{longtable}[]{@{}
  >{\centering\arraybackslash}p{(\linewidth - 6\tabcolsep) * \real{0.1053}}
  >{\centering\arraybackslash}p{(\linewidth - 6\tabcolsep) * \real{0.1317}}
  >{\centering\arraybackslash}p{(\linewidth - 6\tabcolsep) * \real{0.1185}}
  >{\centering\arraybackslash}p{(\linewidth - 6\tabcolsep) * \real{0.1209}}@{}}
\toprule\noalign{}
\begin{minipage}[b]{\linewidth}\centering
\[\mathbf{\alpha}\]
\end{minipage} & \begin{minipage}[b]{\linewidth}\centering
\textbf{RMSE\_\emph{Vm}}
\end{minipage} & \begin{minipage}[b]{\linewidth}\centering
\textbf{RMSE\_\emph{p}}
\end{minipage} & \begin{minipage}[b]{\linewidth}\centering
\textbf{RMSE\_}\(\mathbf{\nu}_{\mathbf{eff}}\)
\end{minipage} \\
\midrule\noalign{}
\endhead
\bottomrule\noalign{}
\endlastfoot
0.5 & 0.025 & 0.005 & 1e-6 \\
1.5 & 0.025 & 0.005 & 1.2e-6 \\
2.5 & 0.024 & 0.005 & 1.5e-6 \\
3.5 & 0.024 & 0.006 & 1.8e-6 \\
4.5 & 0.025 & 0.006 & 2.7e-6 \\
6 & 0.027 & 0.007 & 5e-6 \\
7 & 0.029 & 0.008 & 7.9e-6 \\
8 & 0.031 & 0.009 & 1.3e-5 \\
9 & 0.032 & 0.009 & 1.9e-5 \\
10 & 0.033 & 0.009 & 2.8e-5 \\
\end{longtable}
}

Table 5 shows that the quantitative RMSE results are consistent with the
trends observed in the flow field visualizations. Under interpolation
conditions, corresponding to angles of attack within the training range,
the RMSE values of the velocity magnitude \(V_{m}\), and the pressure p
remain close to each other and exhibit only minor variations with
respect to the angle of attack. This indicates that when the target
conditions lie within the training distribution, the Meta-PINNs maintain
stable and consistent prediction accuracy across different flow
variables.

Under extrapolation conditions, the RMSE values increase monotonically
as the angle of attack rises from 6° to 10°. The growth of errors
reflects the increasing difficulty of prediction as the operating
condition moves further away from the training range. While the RMSE
values of the velocity components and pressure increase gradually, the
effective viscosity exhibits a more pronounced rise in error magnitude,
indicating a higher sensitivity to extrapolation. This behavior is
consistent with the fact that the effective viscosity is closely related
to turbulence production and transport processes, which become more
complex at higher angles of attack.

Overall, despite the gradual increase in error levels under
extrapolation conditions, the magnitude of the RMSE remains controlled
for all variables, demonstrating that the Meta-PINNs retain robust
predictive capability even when applied to flow conditions beyond the
training range.

\begin{figure}[htbp]
\centering
\includegraphics[width=0.8\textwidth]{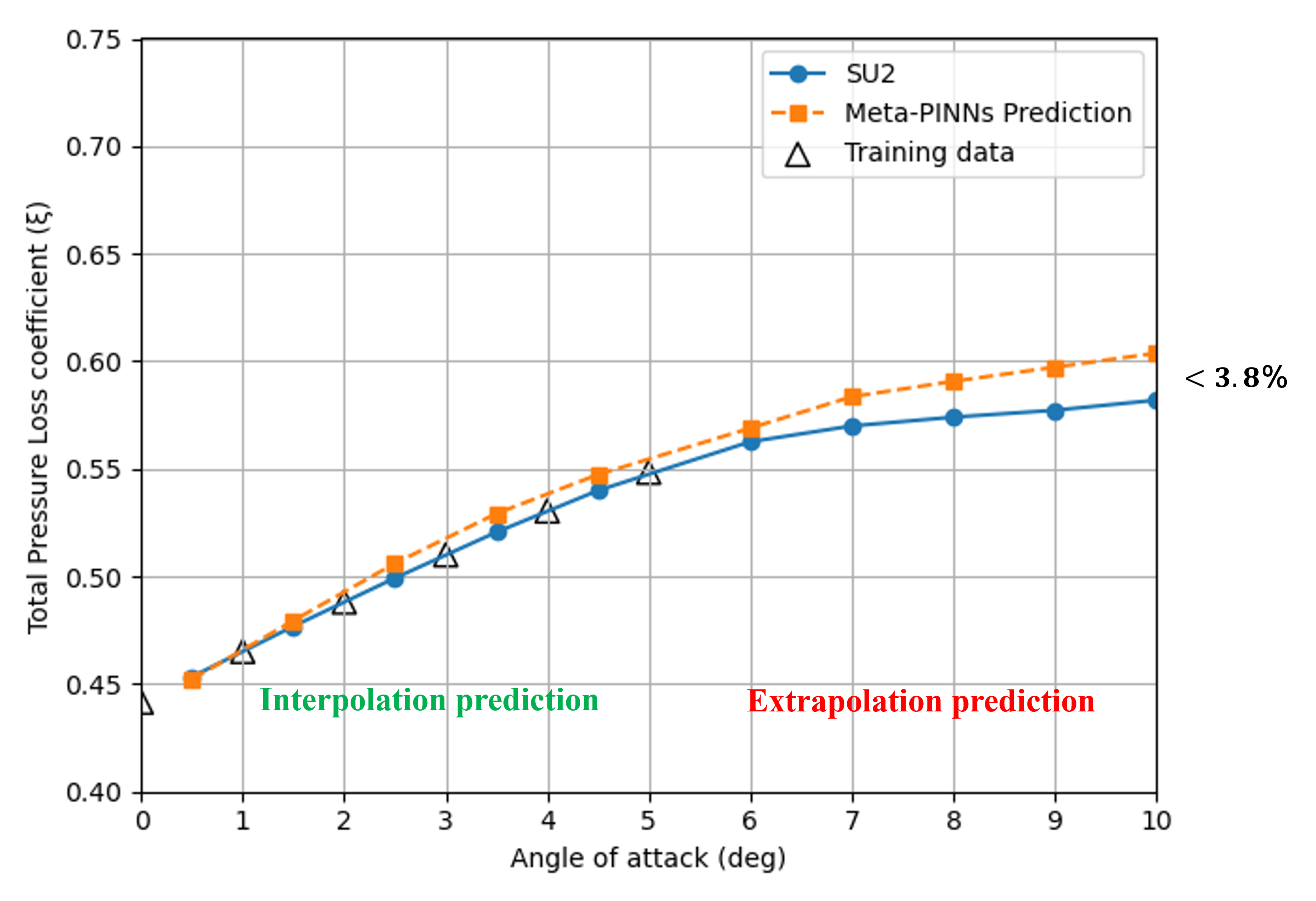}
\begin{flushleft}
\textbf{Fig. 9} \textbf{Total pressure loss coefficient comparison
between CFD results and prediction from Meta-PINNs}
\end{flushleft}

\end{figure}

Figure 9 shows the distribution of total pressure loss coefficient
\(\xi\) with respect to the angle of attack. The total pressure loss
coefficient characterizes the aerodynamic loss within the compressor
cascade by quantifying the decrease in total pressure as the flow
traverses the passage. It is defined as:

\begin{equation}
\xi = \frac{p_{t,\infty} - p_{t,out}}{\frac{1}{2}\rho U_{\infty}^{2}}
\end{equation}

where \(p_{t,\infty}\) is the inlet total pressure, \(p_{t,out}\) is the
total pressure at the outlet surface, \(U_{\infty}\) is the inlet
free-stream velocity. A higher value of \(\xi\) indicates a greater
aerodynamic loss in the flow passage. The results show that within the
interpolation angle-of-attack range of 0.5° to 5°, the total pressure
loss coefficients predicted by Meta-PINNs closely match the SU2 results.
Both the trend of increasing loss with angle of attack and the
corresponding magnitude are accurately captured, indicating that the
Meta-PINNs can reliably learn the dominant loss mechanisms when the
operating condition remains close to the training range.

In the extrapolation regime, the Meta-PINNs continue to provide reliable
predictions at moderate angles of attack, particularly at 6° and 7°,
where the predicted loss coefficients remain close to the reference
results. As the angle of attack increases further, the deviation between
the predicted and reference values gradually becomes more pronounced.
Although the Meta-PINNs exhibit a tendency to slightly overestimate the
growth rate of total pressure loss at higher angles of attack, they are
still able to follow the overall nonlinear evolution of the loss with
respect to the angle of attack. Overall, despite a reduction in
quantitative accuracy under high-angle-of-attack extrapolation
conditions, the prediction error of the total pressure loss coefficient
remains bounded within 3.8\%, demonstrating that the Meta-PINNs retain a
robust predictive capability even beyond the training range.

\textbf{Table 6 computational cost and prediction accuracy comparison of
Meta-PINNs, PINNs and NNs of}
\(\mathbf{\alpha = 10{^\circ}}\)\textbf{based on normalized flow-field
variables}

{\def\LTcaptype{none} 
\begin{longtable}[]{@{}
  >{\centering\arraybackslash}p{(\linewidth - 6\tabcolsep) * \real{0.1185}}
  >{\centering\arraybackslash}p{(\linewidth - 6\tabcolsep) * \real{0.1185}}
  >{\centering\arraybackslash}p{(\linewidth - 6\tabcolsep) * \real{0.1323}}
  >{\centering\arraybackslash}p{(\linewidth - 6\tabcolsep) * \real{0.1071}}@{}}
\toprule\noalign{}
\begin{minipage}[b]{\linewidth}\centering
\textbf{Re}
\end{minipage} & \begin{minipage}[b]{\linewidth}\centering
\textbf{Meta-PINNs}
\end{minipage} & \begin{minipage}[b]{\linewidth}\centering
\textbf{PINNs}
\end{minipage} & \begin{minipage}[b]{\linewidth}\centering
\textbf{NNs}
\end{minipage} \\
\midrule\noalign{}
\endhead
\bottomrule\noalign{}
\endlastfoot
RMSE\_\emph{u} & \textbf{0.022} & 0.154 & 0.206 \\
RMSE\_\emph{v} & \textbf{0.024} & 0.143 & 0.172 \\
RMSE\_\emph{Vm} & \textbf{0.033} & 0.197 & 0.24 \\
RMSE\_\emph{p} & \textbf{0.009} & 0.106 & 0.122 \\
Time cost & \textbf{8:58:06} & 28:41:25 & 10:03:47 \\
\end{longtable}
}

As shown in Tab. 6, Meta-PINNs achieve the best prediction accuracy
among the three models for the flow prediction at
\(\alpha = 10^{{^\circ}}\). The traditional PINNs perform better than
the NNs but still fall noticeably short of Meta-PINNs. Since the NN does
not incorporate any physical constraints, it exhibits the largest
errors. A similar trend is observed in terms of computational cost.
Meta-PINNs require only a small number of adaptive updates for each new
angle of attack, resulting in the shortest runtime of per task. In
contrast, the PINNs must be trained from scratch for every operating
condition, making it the most time-consuming method. The NNs are faster
than the PINNs, but because it still requires retraining for each angle
of attack, its overall efficiency remains lower than that of Meta-PINNs.
The results demonstrate the Meta-PINNs reduce the computational cost by
approximately 45.6\% and 10.8\% relative to the PINNs and NNs,
respectively.

\vspace{0.5em}

\textbf{5 CONCLUSIONS}

This study proposes a novel Meta-PINNs-based flow-field prediction
framework designed to address flow predictions under varying operating
conditions. The framework integrates physical constraints with a
meta-learning strategy, overcoming the inherent limitations of
conventional PINNs. Two representative flow configurations are
investigated: unsteady circular cylinder flow with varying Reynolds
numbers, and compressor cascade flows under different angles of attack,
covering both laminar and turbulent regimes. The main conclusions can be
drawn:

\begin{enumerate}
\def\labelenumi{\arabic{enumi})}
\item
  For the cylinder flow case, Meta-PINNs accurately reproduce the Kármán
  vortex-shedding cycle, wake dynamics, and pressure distribution,
  achieving the highest prediction accuracy at the lowest computational
  cost compared with vanilla PINNs and standard NNs. In terms of
  prediction error, Meta-PINNs improve the prediction accuracy by 1-2
  orders of magnitude relative to PINNs and NNs. Meanwhile, the
  accompanying computational cost is reduced by approximately 95.7 \%
  and 92.1 \% compared with PINNs and NNs, respectively.
\item
  For the compressor cascade flow, Meta-PINNs successfully predict
  velocity vectors, surface pressure distributions, effective viscosity
  fields and downstream wake structures across a broad range of angle of
  attack. The total pressure loss coefficient agrees well with the
  reference data at small and moderate angles of attack. Although
  deviations occur at a higher angle of attack, with a maximum
  prediction error of less than 3.8 \%, Meta-PINNs still reproduce the
  correct overall nonlinear increasing trend. Meta-PINNs are still
  superior to PINNs and NNs in terms of prediction accuracy and
  computing cost. The results demonstrate the Meta-PINNs reduce the
  computational cost by approximately 45.6 \% and 10.8 \% relative to
  the PINNs and NNs, respectively.
\item
  The meta-learned parameter initialization strategy enables rapid
  adaptation to new operating conditions. Thus, Meta-PINNs achieve high
  prediction accuracy for interpolation tasks and maintain reasonable
  accuracy even for extrapolation cases involving substantial parametric
  variations. These findings demonstrate the potential of meta-learning
  to enhance the efficiency and transferability of physics-informed
  models, making the approach well suited for real-world engineering
  applications that require fast, cross-condition flow predictions.
\end{enumerate}

Overall, the Meta-PINNs framework exhibits strong adaptability and
generalization capability for aerodynamic problems with significant
parametric variations. The findings illustrate the potential of
meta-learning to enhance the efficiency and transferability of
physics-informed models, making it well suited for engineering
applications requiring fast, cross-condition flow predictions. Although
the present study focuses on two-dimensional benchmark cases, it
represents an important first step toward more realistic turbomachinery
applications. Future work will extend the proposed Meta-PINNs framework
to three-dimensional flow configurations and more complex industrial
scale problems.

\vspace{0.5em}

\textbf{Conflict of Interest}

There are no conflicts of interest

\vspace{0.5em}

\textbf{Data Availability Statement}

The datasets generated and supporting the findings of this article are
obtainable from the corresponding author upon reasonable request.

\vspace{0.5em}

\textbf{NOMENCLATURE}
\setlength\LTleft{0pt}
\setlength\LTright{0pt}
{\def\LTcaptype{none} 
\begin{longtable}[]{@{}
  >{\raggedright\arraybackslash}p{(\linewidth - 2\tabcolsep) * \real{0.1178}}
  >{\raggedright\arraybackslash}p{(\linewidth - 2\tabcolsep) * \real{0.3564}}@{}}
\endhead
\endlastfoot
$\mathbf{u}$ & Velocity Vector \\
\(u\) & Axial Velocity Component \\
\(v\) & Vertical Velocity Component \\
\(V_{m}\) & Velocity Magnitude \\
\(p\) & Static Pressure \\
\(\mathbf{S}\) & Strain-Rate Tensor \\
\(\nu\) & Molecular Viscosity \\
\(\nu_{eff}\) & Effective Viscosity \\
\(\nu_{t}\) & Turbulent Viscosity \\
\(S_{t}\) & Strouhal Number \\
\(f\) & Vortex-Shedding Frequency \\
\(D\) & Characteristic Length \\
\(U_{\infty}\) & Free-Stream Velocity \\
\(C_{p}\) & Pressure Coefficients \\
\(p_{t}\) & Total Pressure \\
\(p_{t,out}\) & Outlet Total Pressure \\
\(p_{t,\infty}\) & Inlet Total Pressure \\
\(p_{\infty}\) & Free-Stream Static Pressure \\
\(\rho\) & Fluid Density \\
\(x\) & Axial Coordinates \\
\(y\) & Vertical Coordinates \\
\(t\) & Time \\
\(Re\) & Reynolds Number \\
\(\alpha\) & Angle of Attack \\
\(\theta\) & Trainable Parameters \\
\(\theta^{*}\) & Initial Model Parameters \\
\(L_{phys}(\theta)\) & Physics-Informed Loss \\
\(L_{data(\theta)}\) & Data Loss \\
\(L_{meta(\theta)}\) & Meta-PINNs' Residual Loss \\
\(L_{PINN}\) & Total Training Loss \\
\(\mathcal{R}_{mom}\) & Residuals of Momentum Equations \\
\(\mathcal{R}_{cont}\) & Residuals of Continuity Equations \\
\(\omega_{d}\), \(\omega_{\text{p}}\) & Adaptive Weighting
Coefficients \\
\(\mathcal{T}_{i}\) & Prediction Task \\
\(\mathcal{D}_{i}^{\text{sup}}\) & Support Set \\
\(\mathcal{D}_{i}^{\text{qry}}\) & Query Set \\
\(lr\) & Inner Loop Learning Rate \\
\(\theta_{\dot{i}}'\) & Trained Parameters for Each Case \\
\(\beta\) & Outer Loop Learning Rate \\
\(T\) & Full Vortex-Shedding Cycle \\
\(\xi\) & Total Pressure Loss Coefficient \\
CFD & Computational Fluid Dynamic \\
Meta-PINNs & Meta-Learning Enhanced Physics-Informed Neural Networks \\
NNs & Neural Networks \\
PDEs & Partial Differential Equations \\
PINNs & Physics-Informed Neural Networks \\
SciML & Scientific Machine Learning \\
\end{longtable}
}

\vspace{0.5em}

\textbf{REFERENCES}

\begin{enumerate}
\def\labelenumi{\arabic{enumi}.}
\item
  \protect\phantomsection\label{_Ref214289851}{}Li, Z. and Zheng, X.,
  2017. Review of design optimization methods for turbomachinery
  aerodynamics.~\emph{Progress in Aerospace Sciences},~\emph{93},
  pp.1-23.
\item
  \protect\phantomsection\label{_Ref214290103}{}Moukalled, F., Mangani,
  L. and Darwish, M., 2016. \emph{The finite volume method in
  computational fluid dynamics}. Springer.
\item
  \protect\phantomsection\label{_Ref214290187}{}Löhner, R.,
  2008.~\emph{Applied computational fluid dynamics techniques: an
  introduction based on finite element methods}. John Wiley \& Sons.
\item
  \protect\phantomsection\label{_Ref214290356}{}Li, Z., Montomoli, F.,
  Casari, N. and Pinelli, M., 2023. High-dimensional uncertainty
  quantification of high-pressure turbine vane based on multifidelity
  deep neural networks.~\emph{Journal of
  Turbomachinery},~\emph{145}(11), p.111009.
\item
  \protect\phantomsection\label{_Ref214290454}{}Fukami, K., Fukagata, K.
  and Taira, K., 2019. Super-resolution reconstruction of turbulent
  flows with machine learning.~\emph{Journal of Fluid
  Mechanics},~\emph{870}, pp.106-120.
\item
  \protect\phantomsection\label{_Ref214290483}{}Li, Z. and Montomoli,
  F., 2025. Physics-Guided Graph Neural Networks: Numerical
  Investigation of High-Pressure Turbine Flow.~\emph{Journal of
  Turbomachinery},~\emph{147}(12), p.121005.
\item
  \protect\phantomsection\label{_Ref214290536}{}Wu, H., Luo, H., Wang,
  H., Wang, J. and Long, M., 2024. Transolver: A fast transformer solver
  for PDEs on general geometries.~\emph{arXiv preprint
  arXiv:2402.02366}.
\item
  \protect\phantomsection\label{_Ref214290986}{}Lino, M., Fotiadis, S.,
  Bharath, A.A. and Cantwell, C.D., 2023. Current and emerging
  deep-learning methods for the simulation of fluid
  dynamics.~\emph{Proceedings of the Royal Society A},~\emph{479}(2275),
  p.20230058.
\item
  \protect\phantomsection\label{_Ref214291053}{}Cai, S., Mao, Z., Wang,
  Z., Yin, M. and Karniadakis, G.E., 2021. Physics-informed neural
  networks (PINNs) for fluid mechanics: A review.~\emph{Acta Mechanica
  Sinica},~\emph{37}(12), pp.1727-1738.
\item
  Li, Z., Montomoli, F. and Sharma, S., 2024. Investigation of
  compressor cascade flow using physics-informed neural networks with
  adaptive learning strategy.~\emph{AIAA Journal},~\emph{62}(4),
  pp.1400-1410.
\item
  Hanrahan, S.K., Kozul, M. and Sandberg, R.D., 2025. Data Assimilation
  of Transitional and Separated Turbomachinery Flows with
  Physics-Informed Neural Networks.~\emph{Journal of
  Turbomachinery},~\emph{147}(11), p.111011.
\item
  \protect\phantomsection\label{_Ref214291241}{}Zhong, L., Wu, B. and
  Wang, Y., 2023. Accelerating physics-informed neural network based 1D
  arc simulation by meta learning.~\emph{Journal of Physics D: Applied
  Physics},~\emph{56}(7), p.074006.
\item
  \protect\phantomsection\label{_Ref214291251}{}Cheng, S. and
  Alkhalifah, T., 2024. Meta-PINN: Meta learning for improved neural
  network wavefield solutions.~\emph{arXiv preprint arXiv:2401.11502}.
\item
  Toloubidokhti, M., Ye, Y., Missel, R., Jiang, X., Kumar, N., Shrestha,
  R. and Wang, L., 2023. Dats: Difficulty-aware task sampler for
  meta-learning physics-informed neural networks. In~\emph{The Twelfth
  International Conference on Learning Representations}.
\item
  \protect\phantomsection\label{_Ref214291261}{}Wong, J.C., Ooi, C.C.,
  Gupta, A. and Ong, Y.S., 2022. Learning in sinusoidal spaces with
  physics-informed neural networks.~\emph{IEEE Transactions on
  Artificial Intelligence},~\emph{5}(3), pp.985-1000.
\item
  \protect\phantomsection\label{_Ref215928477}{}Spalart, P. and
  Allmaras, S., 1992, January. A one-equation turbulence model for
  aerodynamic flows. In \emph{30th Aerospace Sciences Meeting and
  Exhibit}, pp. 439.
\item
  \protect\phantomsection\label{_Ref215928589}{}Economon, T.D.,
  Palacios, F., Copeland, S.R., Lukaczyk, T.W. and Alonso, J.J., 2016.
  SU2: An open-source suite for multiphysics simulation and design.
  \emph{AIAA Journal}, 54(3), pp.828-846.
\item
  \protect\phantomsection\label{_Ref215931448}{}Williamson, C.H., 1988.
  The existence of two stages in the transition to three-dimensionality
  of a cylinder wake.
\item
  \protect\phantomsection\label{_Ref216015613}{}Ma, W., Ottavy, X., Lu,
  L., Leboeuf, F. and Gao, F., 2011.~Experimental investigations of
  corner stall in a linear compressor cascade. In \emph{ASME 2011 Turbo
  Expo: Turbine Technical Conference and Exposition},~54679, pp. 39-51.
\item
  \protect\phantomsection\label{_Ref216030207}{}Kingma, D.P., 2014.
  Adam: A method for stochastic optimization.~\emph{arXiv preprint
  arXiv:1412.6980}.
\end{enumerate}

\end{document}